\documentclass{aa}  
\usepackage{longtable}
\usepackage{graphicx}
\usepackage{txfonts}
\usepackage{booktabs}
\usepackage[colorlinks=true,
            urlcolor=blue,
            linkcolor=blue,
            citecolor=blue]
            {hyperref}
\newcommand{\swift}{\textit{Swift}}
\newcommand{\xmm}{XMM-\textit{Newton}}
\newcommand{\nustar}{NuSTAR}
\newcommand{\iras}{{IRAS\,23226-3843}}
\newcommand{\plm}{$\pm$}
\newcommand{\rb}[1]{\raisebox{1.5ex}[-1.5ex]{#1}}
\newcommand{\Ha}{H$\alpha$}
\newcommand{\Hb}{H$\beta$}

\makeatletter
 \def\hlinewd#1{%
   \noalign{\ifnum0=`}\fi\hrule \@height #1 \futurelet
    \reserved@a\@xhline}
\makeatother

\newcommand{\mcc}[1]{\multicolumn{1}{c}{#1}}

\newcommand{\kms}{km\,s$^{-1}$}
\begin{document} 

    \title{The outburst of the changing-look AGN \iras{} in 2019}

   \author{W. Kollatschny \inst{1}, 
           D. Grupe \inst{2},
           M.L. Parker \inst{3}, 
           M. W. Ochmann \inst{4,1},
           N. Schartel \inst{5},
           E. Romero-Colmenero \inst{6,7},
           H. Winkler \inst{8},
           S. Komossa \inst{9},      
           P. Famula \inst{1},
           M. A. Probst \inst{1},
           M. Santos-Lleo \inst{5}
         }

   \institute{Institut f\"ur Astrophysik und Geophysik, Universit\"at G\"ottingen,
              Friedrich-Hund Platz 1, 37077 G\"ottingen, Germany\\
              \email{wkollat@gwdg.de}
         \and
         Department of Physics, Geology, and Engineering Technology, Northern Kentucky University, Highland Heights, KY 41076, USA
         \and
         Institute of Astronomy, Madingley Road, Cambridge CB3 0HA, UK
         \and
         Astronomisches Institut (AIRUB), Ruhr-Universit\"at Bochum,
         Universit\"atsstrasse 150, 44801 Bochum, Germany
         \and
         ESA, European Space Astronomy Centre (ESAC), 28692
         Villanueva de la Ca{\~nada}, Madrid, Spain
         \and   
         South African Astronomical Observatory, P.O. Box 9, Observatory 7935, Cape Town, South Africa         
         \and
         Southern African Large Telescope (SALT), P.O. Box 9, Observatory 7935, Cape Town, South Africa
         \and
         Department of Physics, University of Johannesburg, PO Box 524, 2006 Auckland Park,
         Johannesburg, South Africa
         \and
         Max-Planck-Insitut f\"ur Radioastronomie, Auf dem H\"ugel 69,
         D-53121 Bonn, Germany
          }

   \date{Received 22 August 2022; Accepted 29 November 2022}
   \authorrunning{Kollatschny et al.}
   \titlerunning{The outburst of \iras}

% \abstract{}{}{}{}{} 
% 5 {} token are mandatory
 
  \abstract
   {}
   {\iras{} has previously been classified as a changing-look active galactic nucleus (AGN) based on observations taken in the 1990s in comparison to X-ray data (\swift{}, \xmm{}, and \nustar{}) and optical spectra taken after a very strong X-ray decline in 2017. In 2019, \swift{} observations revealed a
   strong rebrightening in X-ray and UV fluxes. We aimed to study this outburst in greater detail.}
   {We took follow-up \swift{}, \xmm{}, and \nustar{} observations of \iras{} together with optical spectra (SALT and SAAO 1.9\,m telescope) from
   2019 until 2021.}
   {\iras{} showed a strong X-ray and optical outburst in 2019. It varied in the X-ray continuum by a factor of 5 and in the optical continuum by a factor of
    1.6 within two months. This corresponds to a factor of 3 after correction for the host galaxy contribution. The Balmer and \ion{Fe}{II} emission-line intensities showed comparable variability amplitudes during the outburst in 2019. The \Ha{} emission-line profiles of \iras{} changed from a blue-peaked profile in the years 1997 and 1999 to a broad double-peaked profile in 2017 and 2019. However, there were no major profile variations in the extremely broad double-peaked profiles despite the strong intensity variations in 2019. One year after the outburst, \iras{} changed its optical spectral type and became a Seyfert type 2 object in 2020. Blue outflow components are present in the optical Balmer lines and in the Fe band in the X-rays. A deep broadband \xmm{}/\nustar{} spectrum was taken during \iras{}'s maximum state in 2019. This spectrum is qualitatively very similar to a spectrum taken in 2017, but by
    a factor of 10 higher. The soft X-ray band appears  featureless. The soft excess is well modeled with a Comptonization model. A broadband fit with a power-law continuum, Comptonized soft excess, and Galactic absorption gives a good fit to the combined EPIC-pn and NuSTAR spectrum. In addition, we see a complex and broadened Fe K emission-line profile in the X-rays. The changing-look character in \iras{} is most probably caused by changes in the accretion rate -- based on the short-term variations on timescales of weeks to months.
        } 
{}

   \maketitle
%
%________________________________________________________________

\section{Introduction}

Seyfert 1 galaxies and quasars vary in the optical and X-ray continua on timescales of hours to decades. They show typical root mean square (rms)
variabilities of 10–20 percent and, on occasion, changes by a factor up to 2 in their optical luminosity \citep{rumbaugh18} on timescales of years.
 Some extremely variable active galactic nuclei (AGN) not only show changes in optical luminosity, but also change their optical spectral type. These optical changing-look AGN exhibit transitions from type 1 to type 2 or vice versa on timescales of months to years. Type 1 AGN show both broad and narrow emission lines in their optical spectra, while type 2 AGN only have narrow emission lines. The optical spectral classification can change as a result of a variation in the accretion rate, accretion disk instabilities, or a variation in absorption, amongst others. Some AGN have shown variability amplitudes by a factor of more than 20 in the X-rays \citep[e.g.,][]{grupe01,grupe10}.
AGN that show X-ray spectral variations, that is,\ when a Compton-thick AGN becomes Compton-thin and vice versa, are also called changing-look AGN \citep{guainazzi02}.

To date, a few dozen Seyfert galaxies and quasars have been known to have changed their optical spectral type, for example NGC\,3516 (\citealt{souffrin73}, \citealt{mehdipour22}), NGC\,7603 \citep{tohline76,kollatschny00},  NGC\,4151 \citep{penston84}, Fairall\,9 \citep{kollatschny85}, NGC\,7582 \citep{aretxaga99}, NGC\,2617 \citep{shappee14}, Mrk\,590 \citep{denney14}, HE\,1136-2304  \citep{parker16, zetzl18, kollatschny18}, Mrk\,1018  \citep{mcelroy16,husemann16,kim18,lyu21}, 1ES\,1927+654 (\citealt{trakhtenbrot19}, \citealt{ricci20},    \citealt{laha22}), NGC\,1566 \citep{Oknyansky19, parker19}, and references therein. Further recent findings are based on spectral variations detected in large-scale surveys, such as the Sloan Digital Sky Survey \citep[SDSS; e.g.,][]{komossa08, lamassa15, rumbaugh18, macleod19}, the Catalina Real-time Transient Survey \citep{graham20}, 
or the Wide-field Infrared Survey Explorer \citep{stern18}. However, in most of these new findings, only a few  spectra of individual galaxies have been acquired to confirm their changing-look character. A recent review on changing-look AGN is given by \cite{komossa22}.

\begin{figure}
\centering
\hspace*{-15mm}
\includegraphics[width=12.cm,angle=0]{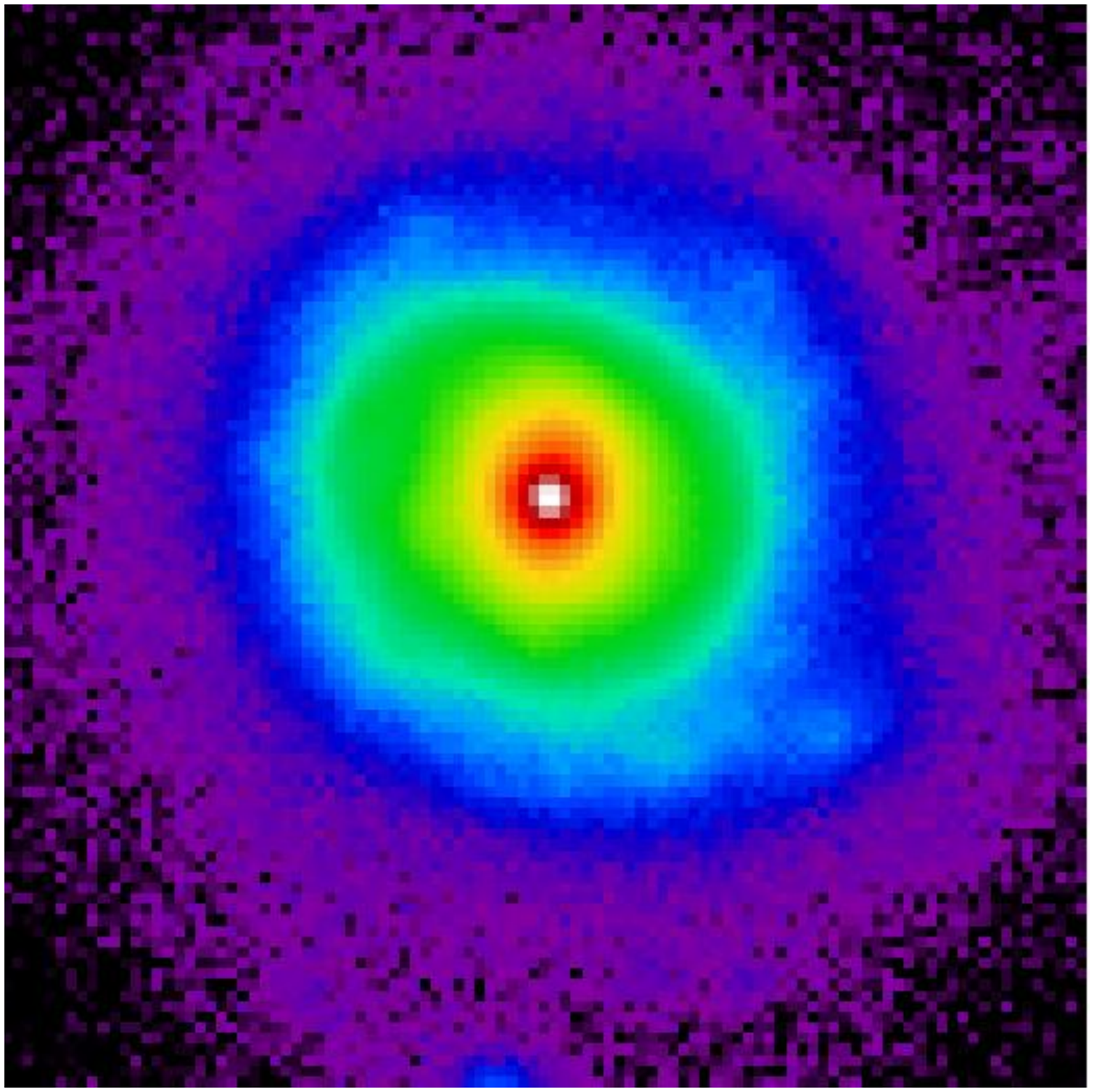}
\caption{DESI legacy image (r band) of \iras{}. North is to the top, and east is to the left. The scale is 0.44 x 0.44 arcmin corresponding to 18.1 x 18.1 kpc.}
 \label{legsurveyIRAS23_r_ds9_3.jpg}
\end{figure}

Here, we discuss an outburst of \iras{} that occurred in 2019. \iras{} is listed as a bright X-ray source in the ROSAT RASS catalog \citep[e.g.,][]{grupe01}. Other designations for \iras{} are CTS\,B11.01, 2MASX\,J23252420-3826492, 6dF J2325242-382649, and HE\,2322-3843. \iras{} is listed as a galaxy in the IRAS Faint Source Catalogue \citep{moshir90}. It has been detected at 60$\mu$m and 100$\mu$m with flux density levels of 0.584\,Jy  and 1.49\,Jy (corresponding to $3.5\times10^{10}$ L$_\odot$). It is a radio source and was listed in the NRAO VLA Sky Survey (NVSS) with a flux density level of 4.3\,mJy \citep{condon98}. \iras{} was classified as a Seyfert type 1 object in an investigation of southern IRAS galaxies in 1991 \citep{allen91}. An apparent magnitude of $13.68\pm 0.10$ was measured for this object in the r band in the Las Campanas Redshift Survey \citep[1991-1996;][]{shectman96}. This
corresponds to an absolute magnitude of M$_{r} = -22.33 \pm 0.51$\,mag. The morphology of \iras{} has been
classified to be of a S0 type based on UKST sky survey images \citep{loveday96}. Figure~\ref{legsurveyIRAS23_r_ds9_3.jpg} displays an r-band image of \iras{} based on the 
Legacy Survey \citep{dey19} indicating a more complex morphology with weak spiral arms.
\\
We discovered a very strong X-ray decline of \iras{} within the \xmm{} slew survey in 2017. Subsequently, we carried out deep follow-up \swift{}, \xmm{}, and \nustar{} observations in combination with optical spectra of \iras{} in 2017 to study its variability behavior. In addition, we reinspected optical, UV, and X-ray data that were taken in the past \citep{kollatschny20}. We found that \iras{} decreased in X-rays by a factor of more than 30
with respect to ROSAT and \swift{} data taken 10 to 27 years earlier. 

In August 2019, we detected a strong increase in X-ray and UV fluxes in \iras{} with \swift{} \citep{grupe19}. Thereupon, we carried out optical follow-up spectroscopy with SALT and with the SAAO 1.9\,m telescope, as well as \xmm\  and \nustar{} observations. Here, we describe and discuss optical and X-ray observations taken between 2019 and 2021. We discuss these results together with findings based on earlier epochs.
Throughout this paper, we assume a lambda cold dark matter ($\Lambda$CDM) cosmology with a Hubble constant of H$_0$~=~73~km s$^{-1}$ Mpc$^{-1}$,  $\Omega_{\Lambda}=0.72$, and $\Omega_{\rm M}=0.28$. With a redshift of {\bfseries $z=0.0359$}, this results in a luminosity distance of $D_{\rm L} = 144$\,Mpc for \iras{} ($\alpha_{2000} =$ 23h 25m 24.2s, $\delta = -38^{\circ}$ 26$^{'}$ 49.2$^{''}$) using the Cosmology Calculator developed by  \cite{wright06}.

\section{Observations and data reduction}
A dramatic increase in the X-ray and UV fluxes of \iras{} was discovered with \swift{} on August 11, 2019. We carried out follow-up studies of \iras{} with \swift{} as well as with \xmm\ and \nustar{} as part of a ToO program to study outbursts of radio-quiet AGN. Furthermore, we took additional optical spectra with SALT as well as with the SAAO 1.9\,m telescope to investigate the optical spectral variations. 

\subsection{\bf X-ray, UV, and optical continuum observations with \swift{}}
\label{sec:swift_observations}

After the discovery of the increasing X-ray flux in 2019 August, we started monitoring \iras{} with the NASA Neil Gehrels Gamma-Ray Burst Explorer Mission \swift{}  \citep{gehrels04}  in X-rays and the UV/optical. All Swift observing dates and exposure times are listed in Table\,\ref{swiftlog}.  All observations in the 0.3-10\,keV band were performed with \swift's X-ray Telescope \citep[XRT;][]{burrows04} in photon counting mode \citep[pc mode;][]{hills05}. Spectra were extracted in a circular region with a radius of 15 pixel (equal to 35$^{\prime\prime}$) for the source region and 100 (equal to 235$^{\prime\prime}$) for the background region using the FTOOL {\it XSELECT}. Using {\it xrtmkarf}, an auxiliary response file (arf) was created and combined with the XRT response file {\it  swxpc0to12s6\_20130101v014.rmf}. 
%{\color{Fussnote noetig?}}
Except for some brighter spectra during the earlier times, generally the spectra were not binned, and analyzed using w-statistics \citep{cash79} as indicated in Table\,\ref{swiftdata}.  This table also lists the count rate, hardness ratio\footnote{Here we define the hardness ratio as $HR = \frac{H-S}{H+S}$ were S are the counts in the 0.3-1.0 keV and H in the 1.0-10.0\,keV bands. Hardness ratios were determined applying  Bayesian statistics with the program {\it BEHR} by \citet{park06}.}, photon index $\Gamma$, the absorption corrected 0.3-10\,keV flux, and the goodness of the fit. 

Typically, observations in the UV-Optical Telescope \citep[UVOT]{roming05} were performed in all 6 photometric filters\footnote{ UVW2 (1928 \AA), UVM2(2246  \AA), UVW1 (2600 \AA), U (3465 \AA), B (4392 \AA), and V (5468 \AA)}. While most observations were performed in the single orbit, in case of multiple orbit observations during one segment, all snapshots were merged using the tool {\it uvotimsum}. The magnitudes in the Vega system and flux densities in each photometric filter were then measured with the tool {\it uvotsource}  using the count rate conversion and calibration, as described in \citet{poole08} and \citet{breeveld10}. Because we applied a source extraction radius of 3", we set the {\it uvotsource} parameter {\it apercorr=curveofgrowth}.  The fluxes and magnitudes listed in Table\,\ref{swiftmag} are corrected for Galactic reddening. We determined the Galactic reddening correction for each filter by applying equation 2 in \citet{roming05} who used the reddening curves published by \citet{cardelli89}. The attenuation in the direction of\,\iras{} is $E_{\rm B-V}=0.021$ \citep{schlafly11}, resulting in the following magnitude corrections: UVW2$_{\rm corr} = 0.20$,  UVM2$_{\rm corr} = 0.244$,
UVW1$_{\rm corr} = 0.177$, U$_{\rm corr} = 0.136$, B$_{\rm corr} = 0.177$, and  V$_{\rm
corr} = 0.083$.

\subsection{{\bf X-ray observations with \xmm\ and \nustar\ }}
The new \xmm\ and \nustar\ observations were taken as part of a ToO program (Proposal ID 084080, PI Schartel) aimed
at catching radio quiet AGN in maximum states \citep[see, e.g.,][]{parker16, parker19}. The observing program consisted of one long \xmm/\nustar\ exposure. In addition, we consider archival \xmm\ and \nustar\ observations
which were taken as part of a ToO program (Proposal ID 76002, PI Schartel) aimed at catching radio quiet AGN in minimum states. The details of these observations are given in Table~\ref{table:xray_observations.}.

\begin{table*}
\centering
\caption{ Details of the \xmm\ and \nustar\ X-ray observations in 2019.}
\label{table:xray_observations}
\begin{tabular}{l c c c c}
\hline  \hline
\noalign{\smallskip}
Mission & Obs. ID &  MJD &  Start date & Exp. time\\
        &         &      &             &   [ks]   \\    
\hline
\xmm\     & 0840800501  &  58795  & 2019-11-07 22:51:54  & 88.29 \\
\nustar\  & 80502607002 &  58795  & 2019-11-07 19:10:11  & 54.83 \\
\hline
\end{tabular}
\tablefoot{
Exposure times are the final clean exposures after filtering for background flares.
}
\label{table:xray_observations.}
\end{table*}

We reduced the \xmm\ EPIC-pn data using the \texttt{epproc} tool from the \xmm\ Science Analysis Software (SAS) version 20.0.0.  We extracted source photons from a 20$^{\prime\prime}$ circular extraction region centered on the source PSF, and background photons from a larger circular region on the same chip, avoiding contaminating sources, detector edges, and the high copper background region of the chip. We also reduced the RGS data; however, we found no evidence for absorption or emission line features in the spectra so we do not consider it further in this work.

We reduced the \nustar\ data using the \nustar\ Data Analysis Software (NuSTARDAS) version 2.1.0, included as part of HEASOFT version 6.29. We extracted source photons from a 30$^{\prime\prime}$ circular extraction region centered on the source PSF, and background photons from a larger circular region on the same chip.

We rebinned all spectra using the \texttt{ftgrouppha} tool, following the optimal binning scheme of \citet{Kaastra16} and with an additional requirement of at least 30 counts per bin. We fit the spectra using \textsc{xspec} version 12.12.0.

\subsection{Optical spectroscopy}
\begin{table}
\centering
\tabcolsep+1.5mm
\caption{Log of spectroscopic observations of \iras{}.} 
%\vspace{3mm}
\begin{tabular}{c c r l l}
\hline \hline
\noalign{\smallskip}
MJD & UT Date & \mcc{Exp. time} & Tel. & Obs. cond.\\
    &   &  \mcc{[s]} &  \\
    
\hline 
50724         &       1997-10-03   &       1800  & SAAO 1.9\,m & clear; 1\farcs{}8  \\

51350         &       1999-06-21      &      600  & Cerro-Tol. 4\,m   &   clouds; --  \\

57883        &       2017-05-10      &      900  & SALT  & clear; 1\farcs{}5  \\

57916         &       2017-06-12      &      900  & SALT & clear; 2\farcs{}0  \\ 

58727         &       2019-09-01      &    2400 & SAAO 1.9\,m   & clear; 1\farcs{}8\\

58736         &       2019-09-10      &     1200  & SALT & clouds; 2\farcs{}0 \\

58750         &       2019-09-24      &      900  & SALT & clear; 1\farcs{}5  \\

58794         &       2019-11-07      &      900  & SALT & clear; 1\farcs{}4  \\

58823         &       2019-12-06      &      900  & SALT & clear; 1\farcs{}3   \\

59053         &       2020-07-23      &    2400 & SAAO 1.9\,m   & clear;  1\farcs{}8 \\
\hline
\vspace{-.7cm}
\end{tabular}
\tablefoot{
The observations have been taken with SALT in 2019 as well as with the SAAO 1.9\,m telescope in 1997, 2019 and 2020. In addition, we give the observing dates already presented in  \cite{kollatschny20}  (Cerro-Tololo 4\,m: 1999; SALT: 2017).  We list the Modified Julian date, the UT date, the exposure time, the telescope, and observing conditions.
}
\label{saltlog}
\end{table}

\subsubsection{Spectroscopy with SALT}
After the detection of the increasing X-ray flux, we took four optical spectra of \iras{} with the Southern African Large Telescope (SALT)  between 2019 September and December. One optical spectrum was secured simultaneously to the deep X-ray observation with \xmm{} on 2019 November 07. The log of our optical spectroscopic observations with SALT is given in Table~\ref{saltlog}. In addition, we present the observing data from 1999 and 2017 \citep{kollatschny20} and three observations taken with the SAAO 1.9\,m telescope (see \ref{sec:SAAO_1.9}).

The spectroscopic observations were taken under identical instrumental conditions with the Robert Stobie Spectrograph using the PG0900 grating. The slit width was fixed to 2\farcs{}0 projected onto the sky at an optimized position angle to minimize differential refraction. Furthermore, all observations were taken at the same air mass thanks to the particular design feature of the SALT. The spectra were taken with exposure times of 15 to 20 minutes (see Table~\ref{saltlog}). Seeing full width at half maximum (FWHM) values ranged between 1.3 to 2\arcsec.

We covered the wavelength range from 4330 to 7376~\AA\  at a spectral resolution of 6.5~\AA\ . The observed wavelength range corresponds to a wavelength range from 4180 to 7120~\AA\ in the rest frame of the galaxy. There are two gaps in the spectrum caused by the gaps between the three CCDs: one between the blue and the central CCD chip, and one between the central and red CCD chip covering the wavelength ranges 5347--5416~\AA\  and 6386--6454~\AA\
(5162--5228~\AA\ and 6165--6230~\AA\ in the rest frame ). All spectra were wavelength corrected to the rest frame of the galaxy ($z=0.0359$). 

In addition to the galaxy spectra, we also observed necessary flat-field and ThAr and Xe arc frames, as well as a spectrophotometric standard star for flux calibration (LTT\,1020). The spatial resolution per binned pixel was
0\farcs{}2534. We extracted eight columns from our object spectrum, corresponding to 2\farcs{}03. The reduction of
the spectra (bias subtraction, cosmic-ray correction, flat-field correction, 2D wavelength calibration, night sky subtraction, and flux calibration) was made in a homogeneous way with IRAF
reduction packages \citep[e.g.,][]{kollatschny01}.

Great care was taken to ensure high-quality intensity and wavelength calibrations to keep the intrinsic measurement errors very low, as described in \citet{kollatschny01,kollatschny03,kollatschny10}. The AGN spectra (and our calibration star spectra) were not always taken under photometric conditions. Therefore, we calibrated the spectra
to the same absolute [\ion{O}{iii}]\,$\lambda$5007 flux of $9.27 \times 10^{-15} \rm erg\,s^{-1}\,cm^{-2}$  taken on 2017 May 10, under clear conditions. The flux of the narrow emission line [\ion{O}{iii}]\,$\lambda$5007 is considered to be constant on timescales of years to decades.

\subsubsection{Spectroscopy with the SAAO 1.9\,m telescope} \label{sec:SAAO_1.9}
The spectra taken on 1997 October 03, 2019 September 01 and 2020 July 23 were obtained with the 1.9\,m telescope at the South African Astronomical Observatory in Sutherland (South Africa). The first spectrum was recorded with the CCD spectrograph in use at the time, whereas we utilized the improved SpUpNIC spectrograph \citep{Crause19} for the 2019 and 2020 observations. All observations used a low resolution grating (300 lines/mm, dispersion of 2.7 \AA/pixel) set at an angle to record the full optical range and using a slit width corresponding to 2\farcs{}7 on the sky. The AGN spectrum  taken in 1997  was extracted over a spatial region of 5\arcsec. The spectra taken in 2019 and 2020 
were extracted over 10\arcsec.
The wavelength was determined by means of observations of the emission spectrum produced by an argon lamp. The flux calibration was achieved through observations on the same night of spectrophotometric standard stars from the compilation of \citet{Hamuy94} (LTT\,9491, LTT\,7379, LTT\,377). We note that due to seeing fluctuations and possible thin cloud presence the flux calibration was later adjusted so that the [\ion{O}{III}] emission line strengths corresponded to those measured in the SALT spectra. Seeing was 2\arcsec or better for all observations, and weather conditions were generally good.

\section{Results}

\subsection{\bf X-ray, UV, and optical continuum band variations based on \swift{}}
\label{sec:swift_results}
\begin{figure*}
\centering
%\vspace*{-20mm} 
\includegraphics[width=17cm,angle=0]{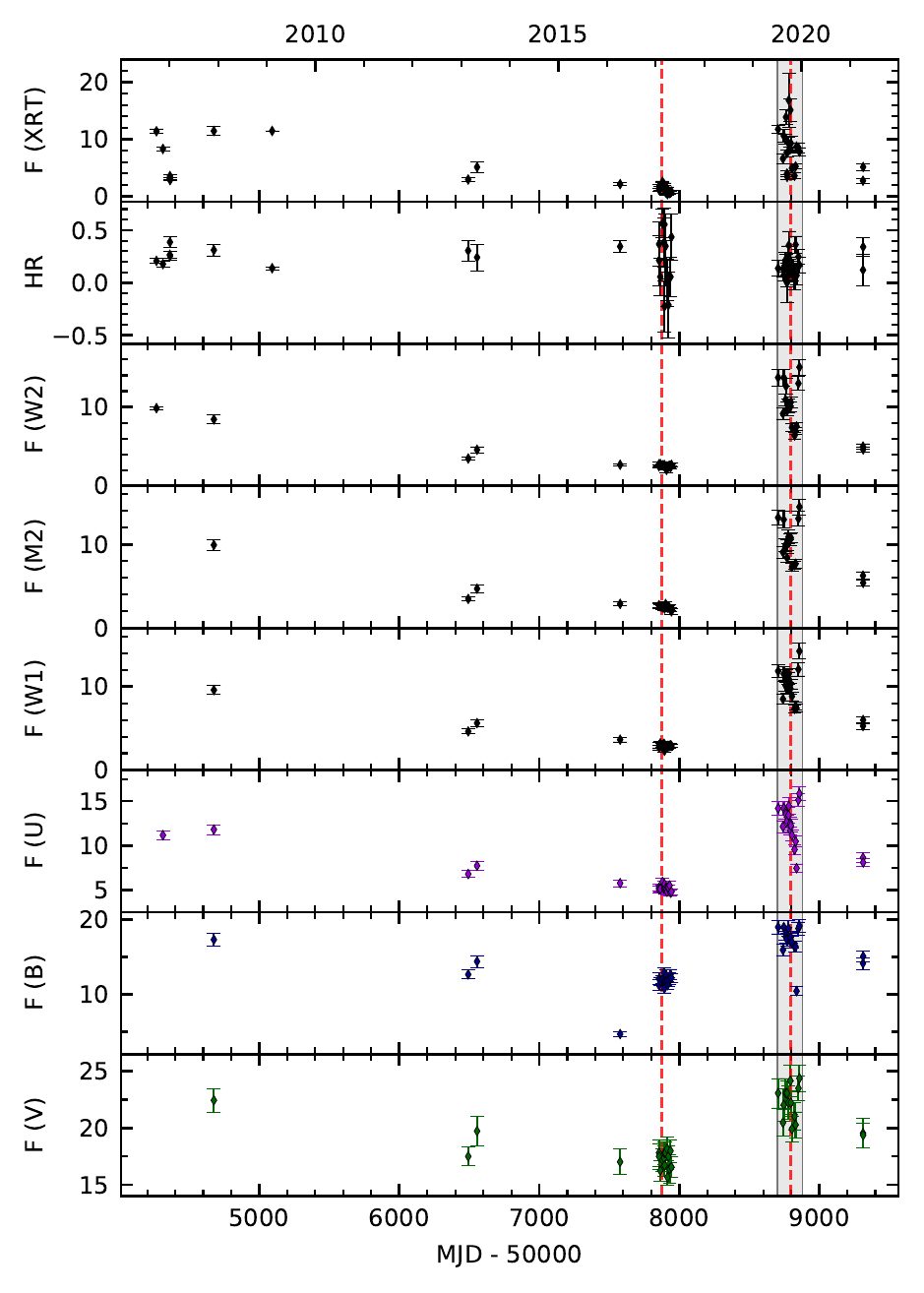}
\vspace*{-5mm} 
\caption{Combined X-ray, UV, and optical light curves taken with the \swift{} satellite for the years 2007 until 2021. The fluxes are given in units of 10$^{-12}$ ergs s$^{-1}$ cm$^{-2}$. HR is the X-ray hardness ratio. The red lines indicate the dates of our deep  \xmm{} observations. The observations for the year 2019 (shaded area) are presented in greater detail in Figure~\ref{fig_dirk_xrt_uvot_lc19.pdf}.}
\label{fig_dirk_xrt_uvot_lc07_19.pdf}
\end{figure*}

The \swift{} 0.3--10\,keV, UV, and optical continuum light curves for the years 2007 to 2021 are shown in Fig.~\ref{fig_dirk_xrt_uvot_lc07_19.pdf} along with the X-ray  and the hardness ratios (see Sect.~\ref{sec:swift_observations}). All measurements are listed in Tables~\ref{swiftdata} and
\ref{swiftmag}.

\begin{figure*}
\centering
%\vspace*{-20mm} 
\includegraphics[width=17.cm,angle=0]{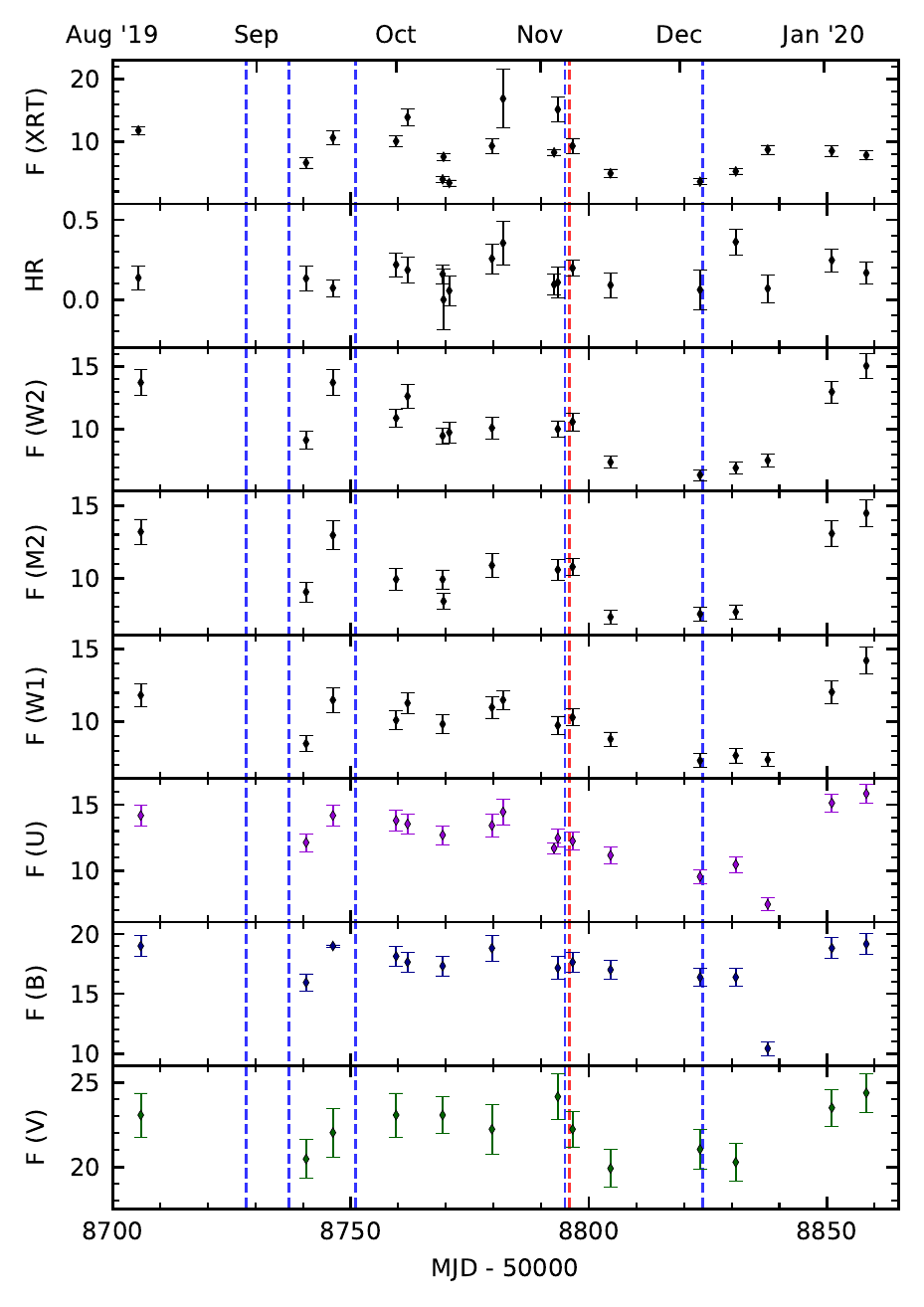}
\vspace*{-5mm} 
\caption{Combined X-ray, UV, and optical light curves
taken with the \swift{} satellite for the year 2019. The fluxes are given in units of 10$^{-12}$ ergs s$^{-1}$ cm$^{-2}$. Furthermore, the dates of our optical spectral observations are indicated as blue lines. We note that the deep XMM-Newton observation was performed on MJD 58795 (red line). }
\label{fig_dirk_xrt_uvot_lc19.pdf}
\end{figure*}

\begin{figure}
    \centering
    \includegraphics[width=\linewidth]{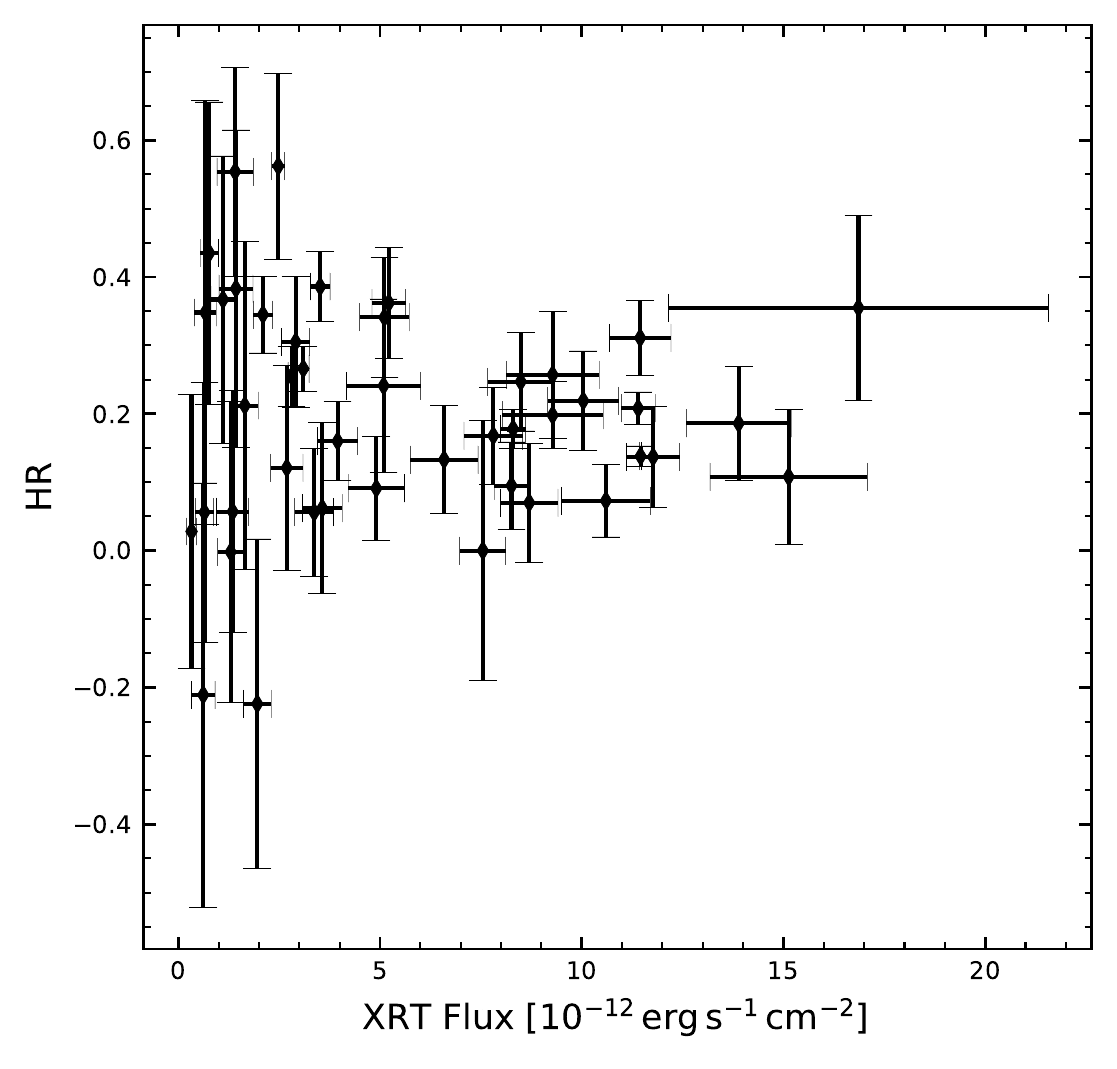}
    \caption{Hardness ratio (HR) of our Swift observations as a function of X-ray flux.  There is no systematic trend of HR with flux.}
    \label{HR_X_flux_20221026.pdf}
\end{figure}

When \swift\ observed  \iras{} as part of our GI filling program on 2019 August 11, it found it in an X-ray high state. The AGN had increased its X-ray flux in the 0.3-10\,keV band by a factor of more than 30 compared with the low-state observation in 2017. On 2019 August 11, \swift\ measured a flux of $1.18\times 10^{-11} $~erg~s$^{-1}$~cm$^{-2}$.  In the UV, the flux increased by about 2 magnitudes as listed in Table~\ref{swiftmag}. Figure~\ref{fig_dirk_xrt_uvot_lc19.pdf} shows the X-ray, UV, and optical \swift{} light curves
for the detailed campaign in 2019. 
The UV and optical \swift{} bands closely follow the 0.3--10\,keV X-ray light curve, which exhibits the strongest variability amplitudes and varies by a factor of more than 50. However, even though we do find changes in the hardness ratio over the years from 2007 until 2021, there is no systematic trend of HR with flux, as illustrated in Fig.~\ref{HR_X_flux_20221026.pdf}.

Table~\ref{swiftvarstatistics} gives the variability statistics based on the \swift{} continua (XRT, W2, M2, W1,
 U, B, and V). 
\begin{table}
\centering
\tabcolsep+1.5mm
\caption{Variability statistics of the \swift{} continua  XRT, W2, M2, W1, U, B, and V.} 
\begin{tabular}{lrrrccc}
\hline \hline
\noalign{\smallskip}
 Band & F$_\text{max}$  & F$_\text{min}$  & R$_\text{max}$  & <F>  &  $\sigma_\text{F}$ &  F$_\text{var}$ \\
\noalign{\smallskip}
\hline 
\noalign{\smallskip}
XRT  & $16.86$ & $0.33$ & $51.09$ & $5.48$ & $4.43$ & $0.79\pm0.09$ \\
W2  & $15.04$ & $2.03$ & $7.41$ & $6.38$ & $4.09$ & $0.63\pm0.07$ \\
M2  & $14.48$ & $1.98$ & $7.31$ & $6.27$ & $4.02$ & $0.64\pm0.08$ \\
W1  & $14.22$ & $2.31$ & $6.16$ & $6.51$ & $3.66$ & $0.56\pm0.07$ \\
U  & $15.84$ & $4.69$ & $3.38$ & $9.03$ & $3.73$ & $0.41\pm0.05$ \\
B  & $19.16$ & $4.68$ & $4.09$ & $14.32$ & $3.3$ & $0.19\pm0.03$ \\
V  & $24.38$ & $15.67$ & $1.56$ & $19.57$ & $2.61$ & $0.12\pm0.02$ \\
\noalign{\smallskip}    
\hline
\end{tabular}
\label{swiftvarstatistics}
\tablefoot{ In units of $10^{-12}$ erg s$^{-1}$ cm$^{-2}$.}
\end{table}
We give the minimum and maximum fluxes F$_{\rm min}$ and F$_{\rm max}$, peak-to-peak amplitudes R$_{\rm max}$ = F$_{\rm max}$/F$_{\rm min}$, the mean flux $<$F$>$ over the period of observations from MJD 54266.18 to 59313.26, the standard deviation $\sigma_{\rm F}$,  and the fractional variation $F_{\rm var}$  as defined by \cite{rodriguez97}. The $F_{\rm var}$ uncertainties are defined in \cite{edelson02}. We observe a clear decrease of the peak-to-peak amplitude and the fractional variation with increasing wavelength.

\subsection{ Optical continuum and spectral line variations}

\begin{figure*}
\centering
\includegraphics[width=17cm,angle=0]{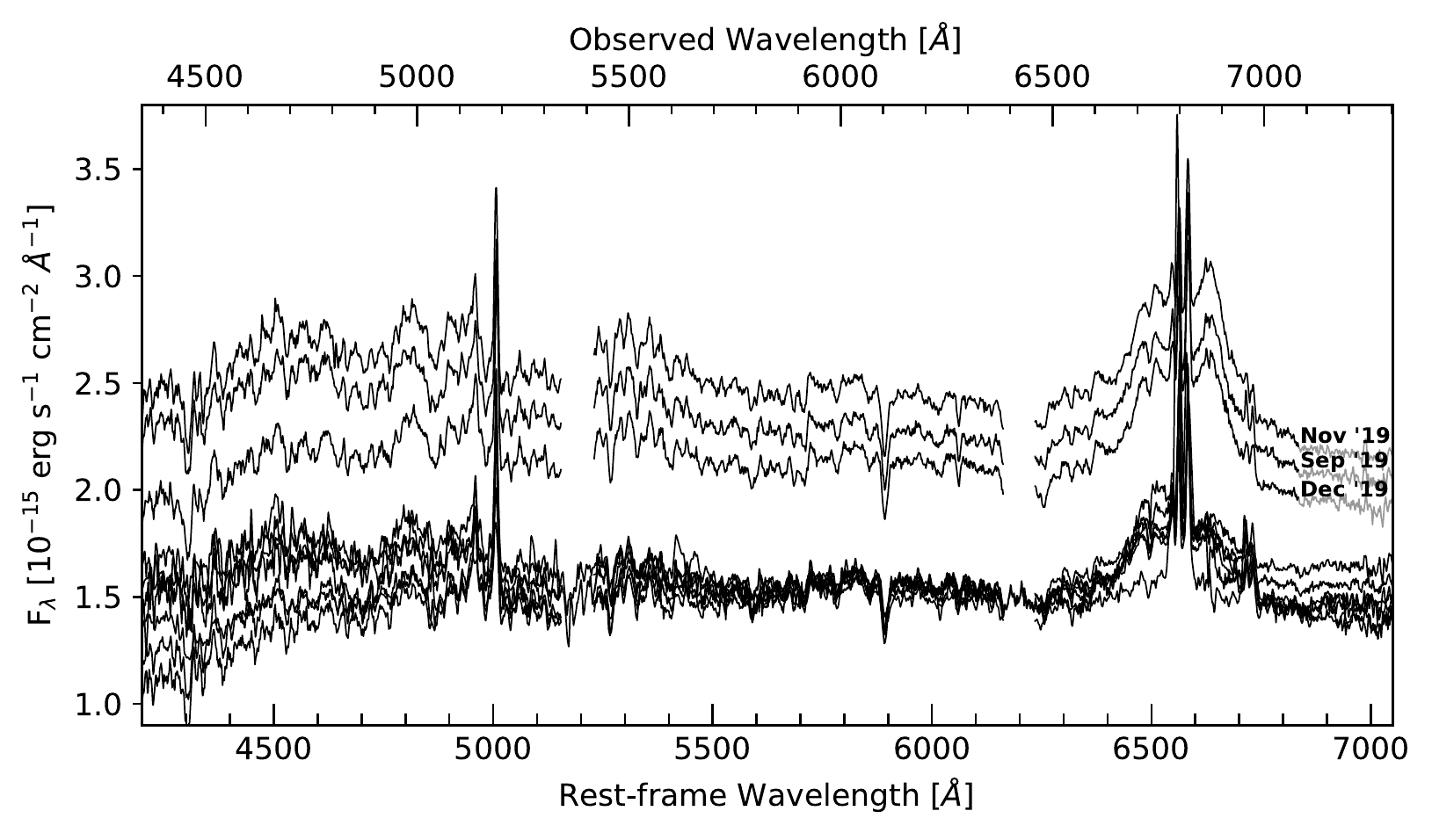}
\vspace*{-3mm} 
\caption{Reduced optical spectra of \iras{} for the years 1997 until 2020. The spectra were calibrated to the same absolute [\ion{O}{iii}]\,$\lambda$5007 flux of $9.27 \times 10^{-15} \rm erg\,s^{-1}\,cm^{-2}$  taken on 2017 May 10, under clear conditions. The three outburst spectra from 2019 are indicated by their observation date.}
\label{IRAS23226_all_spectra_20220326.pdf}
\end{figure*}

The optical spectra were taken with different telescopes, different spectrographs and also under different
atmospheric conditions. For an accurate comparison we therefore had to intercalibrate our spectra. The intercalibration of our spectra shown in Fig.~\ref{IRAS23226_all_spectra_20220326.pdf} was carried out with respect to the [\ion{O}{iii}]\,$\lambda$5007 line as well as with respect to the narrow \Ha{}, [\ion{N}{ii}], and [\ion{S}{ii}] lines as the flux of narrow emission lines is considered to be constant on timescales of years. The two SAAO 1.9\,m spectra that were integrated over a wider spatial region of 10\arcsec{} (2019-09-01 and 2020-07-23) contain a higher contribution from the host galaxy as well as additional non-nuclear line emission. We subtracted a continuum flux of 0.2 $\times$ 10$^{-15}$\,erg\,s$^{-1}$\,cm$^{-2}$\,\AA$^{-1}$ to correct for this additional host galaxy contribution.
A relative flux accuracy of only about 5\% was achieved for [\ion{O}{iii}]$\lambda$\,5007 in these nuclear emission line spectra because the spectra of \iras{} show an additional strong contribution from the underlying host galaxy and the narrow emission line region is slightly extended.

\begin{figure*}
\centering
\includegraphics[width=17cm,angle=0]{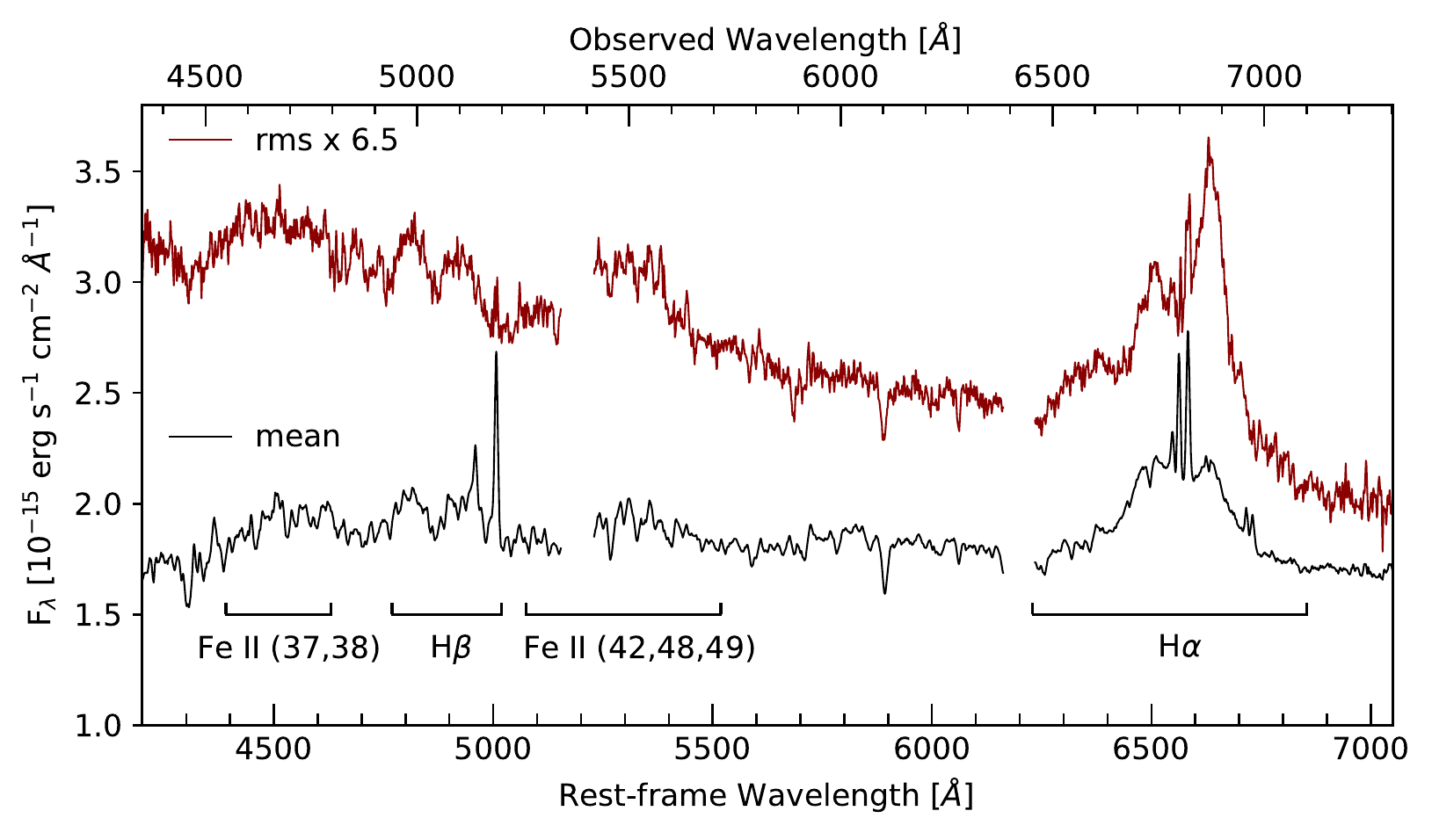}
\caption{Integrated mean (black) and rms (red) spectra of \iras{}. The rms spectrum has been scaled by a factor of 6.5 to enhance  weaker line structures. The Balmer lines \Ha{} and \Hb{} as well as the \ion{Fe}{ii} blends are indicated by black lines.}
\vspace*{-3mm} 
\label{IRAS23226_AVG_RMS_20221026_legend.pdf}
\end{figure*}

We present our reduced optical spectra of \iras{} for the years 1997 to 2020 in Fig.~\ref{IRAS23226_all_spectra_20220326.pdf}. The spectra show a strong contribution from the host galaxy with a clear stellar absorption imprint. The observed continuum fluxes at $\sim 5500-6000$\,\AA{}  do not vary much between the epochs, except for the outburst spectra from 2019 September to 2019 December. However, the continuum gradient shows slight variations with a varying strength of the continuum slope.
The mean and rms spectra of \iras{} are shown in Fig.~\ref{IRAS23226_AVG_RMS_20221026_legend.pdf}. They are based on all spectra except for the two SAAO 1.9\,m spectra that were integrated over an extended  region of 10\arcsec. The rms spectrum that presents the variable part of the spectra, that is to say the variable part of the emission lines \ in particular, was scaled by a factor of 6.5 to allow for a better comparison with the mean spectrum, and  to enhance weaker line structures. 
%
%------------------------------------------------------------------------------
%
%
%
%---------------------------------------------------------------------------
%                                                

 %---------------------------------------------------------------------------

%

The broad emission line spectrum is dominated by broad double-peaked \Hb{} and \Ha{} emission lines, as well as permitted broad \ion{Fe}{ii} blends of the multiplets 37, 38  and 42, 48, 49. These broad \Ha{} and \Hb{} lines as well as the \ion{Fe}{ii} blends are highly variable as can be seen in the rms spectrum in Fig.~\ref{IRAS23226_AVG_RMS_20221026_legend.pdf}. In addition to the broad lines, the typical narrow AGN emission lines of [\ion{O}{iii}]$\lambda\lambda\, 4595,5007$, [\ion{O}{i}]$\lambda\, 6300$, [\ion{N}{ii}]$\lambda\lambda\, 6548,6584$, [\ion{S}{ii}]$\lambda\lambda\, 6716,6731$, and narrow \Ha{} are present. Before determining the emission line intensities, we subtracted the strong contribution of the host galaxy, as well as the underlying power-law continuum by means of a spectral synthesis. The details of this synthesis are outlined in Sect.~\ref{sec:host_galaxy}. The wavelength ranges of the broad emission lines, of the continua as well as of the pseudo-continua we used to subtract a continuum below the broad Balmer lines and the \ion{Fe}{ii} blends are given in Table~\ref{wavelength-ranges}. 

\begin{table}
\centering
\caption{Wavelength ranges of the (pseudo-)continua, the Balmer lines, and \ion{Fe}{ii} blends in the rest-frame.}
\label{wavelength-ranges}
\begin{tabular}{lcc}
\hline \hline
Continuum, Line & Wavelength range \\
           &    [\AA]         \\
\hline
Cont 4230    & 4210 -- 4245   \\
Cont 5040    & 5022 -- 5056    \\
Cont 5545    & 5520 -- 5570   \\
Cont 5965    & 5920 -- 6010    \\
Cont 7020    & 6990 -- 7050    \\
\hline
\ion{Fe}{ii}(37,38)  & 4390 -- 4630   \\
H$\beta_{\rm broad}$ & 4770 -- 5019   \\  
\ion{Fe}{ii}(42,48,49)   & 5057 -- 5519   \\
H$\alpha_{\rm broad}$ & 6230 -- 6850  \\  
H$\alpha_{\rm inner}$ & 6439 -- 6776  \\  
\hline
\end{tabular}
\tablefoot{
 H$\alpha_{\rm inner}$ has the same extent in velocity space as H$\beta_{\rm broad}$.
 }
\end{table}
We list in Tab.~\ref{BLR_variability} the extinction corrected observed continua fluxes at $4230\pm20$\,\AA{} and $5040\pm20$\,\AA{},  as well as the broad-line fluxes of \Ha{} and \Hb.  For \Ha{}, we also give the flux of the inner region (H$\alpha_{\rm inner}$) which has the same extent in the velocity space as H$\beta_{\rm broad}$ ($-5640$\,\kms{} to $+7800$\,\kms{}). In addition, we list the flux of the \ion{Fe}{ii} blend (42, 48, and 49). Here, we linearly interpolated the flux for the gap between 6165 and 6230\,\AA{}. The broad H$\beta_{\rm broad}$ line has been corrected for the contribution of the narrow line fluxes of  H$\beta_{\rm narrow}$,  [\ion{O}{iii}]$\lambda\,$4959, and [\ion{O}{iii}]$\lambda\,$5007. 
The broad H$\alpha_{\rm broad}$ line as well as the inner H$\alpha_{\rm inner}$ line have been corrected for the contribution of the narrow line fluxes of H$\alpha_{\rm narrow}$, [\ion{O}{i}]$\lambda$\,6300, [\ion{N}{ii}]$\lambda\lambda\,6548, 6584$, and [\ion{S}{ii}]$\lambda\lambda$\,6716, 6731.
%(15.9 10$^{-15}$\,erg\,s$^{-1}$\,cm$^{-2}$ for all
%epochs except for 1997-10-03: 24.3 10$^{-15}$\,erg\,s$^{-1}$\,cm$^{-2}$, and %2019-09-01: 27.4 10$^{-15}$\,erg\,s$^{-1}$\,cm$^{-2}$). 
In addition, we present the Balmer decrement \Ha{}/\Hb{} based on the inner
H$\alpha_{\rm inner}$ region ($-5640$\,\kms up to $+7800$\,\kms{}). The flux of the \ion{Fe}{ii}(37,38) blend exhibits the same trend of showing a maximum at 2019-11-07 similar to the \ion{Fe}{ii}(42,48,49) blend. However, the \ion{Fe}{ii}(37,38) blend holds a major error caused by the extrapolation of the spectral synthesis and is therefore not listed. 
\begin{table*}
\caption{Fluxes of the optical continua at $4230\pm20$\AA{} and $5040\pm20$\AA{}, 
the broad-line intensities of \Ha{} and \Hb{} after subtraction of the narrow line components, 
the flux of the \ion{Fe}{ii} blend (42, 48, and 49), and the Balmer decrement \Ha/\Hb{} for the
inner line region. } 
\label{BLR_variability}
\begin{tabular}{l c c  c c c c c}
\hline \hline
UT-Date & Cont$_{4230}$ & Cont$_{5040}$ & \Hb{} & \ion{Fe}{ii}(42,48,49) & \Ha{} &  H$\alpha_{\rm inner}$ &  (\Ha{} / \Hb{})$_{\rm inner}$\\ 
\hline
1997-10-03      &       1.54    $\pm$   0.29    &       1.53    $\pm$   0.29    &  23.7   $\pm$   4.0     &       17.6    $\pm$   8.0     &       104.0   $\pm$   17.0    &       92.9    $\pm$   16.0    &       3.9 $\pm$   1.0     \\
1999-06-21      &       1.50    $\pm$   0.23    &       1.51    $\pm$   0.23    &  24.2   $\pm$   3.5     &       28.6    $\pm$   8.0     &       100.5   $\pm$   14.0    &       89.4    $\pm$   13.0    &       3.7 $\pm$   0.8     \\
2017-05-11      &       1.11    $\pm$   0.12    &       1.42    $\pm$   0.15    &  13.2   $\pm$   3.0     &       10.1    $\pm$   6.0     &       93.8    $\pm$   13.0    &       73.9    $\pm$   12.0    &       5.6 $\pm$   1.6     \\
2017-06-13      &       1.24    $\pm$   0.13    &       1.47    $\pm$   0.15    &  13.2   $\pm$   3.0     &       7.7         $\pm$       6.0     &       96.4    $\pm$   13.0    &       75.0    $\pm$   12.0    &       5.7 $\pm$   1.6     \\
2019-09-01      &       1.58    $\pm$   0.27    &       1.57    $\pm$   0.27    &  13.1   $\pm$   4.0     &       12.5    $\pm$   7.0     &       86.1    $\pm$   19.0    &       70.9    $\pm$   17.0    &       3.7 $\pm$   2.2     \\
2019-09-10      &       1.58    $\pm$   0.39    &       1.60    $\pm$   0.40    &  11.8   $\pm$   6.0     &       31.7    $\pm$   12.0 &  115.1   $\pm$   27.0    &       95.8    $\pm$   24.0    &       8.1 $\pm$   4.7     \\
2019-09-24      &       2.32    $\pm$   0.14    &       2.33    $\pm$   0.14    &  26.3   $\pm$   2.0     &       44.6    $\pm$   5.0     &       181.0   $\pm$   9.0         &    148.0   $\pm$   8.0         &   5.6 $\pm$   0.6 \\
2019-11-07      &       2.46    $\pm$   0.15    &       2.51    $\pm$   0.16    &  31.0   $\pm$   2.0     &       50.2    $\pm$   5.0     &       199.6   $\pm$   9.0         &    164.0   $\pm$   8.0         &   5.3 $\pm$   0.5 \\
2019-12-06      &       1.96    $\pm$   0.12    &       2.12    $\pm$   0.13    &  14.7   $\pm$   2.0     &       35.4    $\pm$   5.0     &       155.8   $\pm$   9.0         &    134.0   $\pm$   8.0     &       9.1 $\pm$   1.4 \\
2020-07-23      &       1.39    $\pm$   0.20    &       1.46    $\pm$   0.20    &         5.6 $\pm$       6.0 &    2.5    $\pm$   4.0     &       18.1    $\pm$   20.0         &        15.1   $\pm$   19.0    &  --   \\
\hline
mean values & 1.66 $\pm$ 0.07  & 1.75 $\pm$ 0.04 & 17.7 $\pm$ 2.7  & 24.1 $\pm$ 3.1  & 115.0 $\pm$ 8. & 95.9  $\pm$ 8.  & 5.7 $\pm$   1.1 \\                
\hline
ratio(11-19/06-17)  & 1.98   &  1.71  & 2.35   &  6.52    &  2.07   & 2.19 & 0.93  \\  
\hline
\end{tabular}
\tablefoot{
Continuum flux densities in units of 10$^{-15}$\,erg\,s$^{-1}$\,cm$^{-2}$\,\AA$^{-1}$. Line fluxes in units 10$^{-15}$\,erg\,s$^{-1}$\,cm$^{-2}$.
}
\end{table*} 

  \begin{figure*}
 \centering
   \includegraphics[width=1.0\linewidth,angle=0]{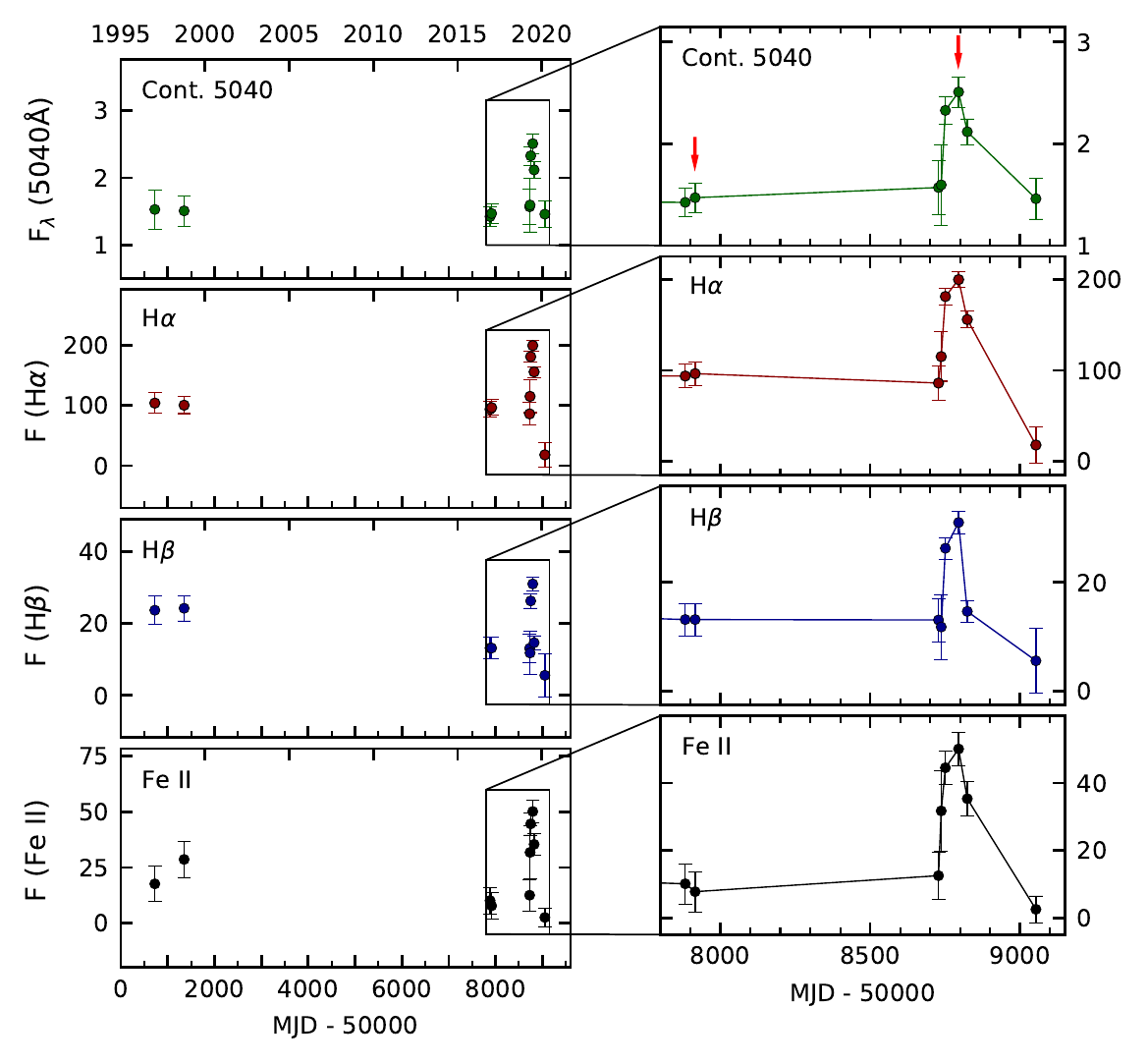}
      \caption{Long-term light curves of the continuum flux density at 5040\,\AA{} (in units of 10$^{-15}$\,erg\,s$^{-1}$\,cm$^{-2}$\,\AA$^{-1}$), as well as of the line
      fluxes of \Ha, \Hb, and \ion{Fe}{ii}(42,48,49) (in units 10$^{-15}$\,erg\,s$^{-1}$\,cm$^{-2}$) for the years 1997 until 2020. The right panel shows, in addition to the observations from 2017 and 2020 the variations in 2019 in more detail. The epochs of the deep \xmm{} observations are indicated by a red arrow.}
       \vspace*{-3mm} 
         \label{Ochmlightcurves_IRAS23226_20220329.pdf}
   \end{figure*}
%  
   %---------------------------------------------------------------------------

We show the long-term light curves of the
continuum flux density at 5040\,\AA{}, as well as of the line fluxes of 
\Ha{}, \Hb{}, and \ion{Fe}{ii}(42,48,49) for the years 1997 until 2020 in Fig.~\ref{Ochmlightcurves_IRAS23226_20220329.pdf}. A second plot gives the
variations in 2019 in more detail. The epochs of the deep XMM observations recorded in parallel at 2017 June 11 and 2019 November 07 are indicated by a red arrow.

\subsection{\bf Host galaxy contribution determined by spectral synthesis}\label{sec:host_galaxy}

We carried out a spectral synthesis of  the galaxy spectra using the Penalized Pixel-Fitting method (pPXF) \citep{cappellari04,cappellari17}. This software extracts the stellar population from absorption-line spectra of galaxies, using a maximum penalized likelihood approach. The host galaxy spectrum of \iras{} is highly contaminated by strong emission line blends. Therefore, we restricted our synthesis to those wavelength ranges that were (nearly) free from emission lines. These regions correspond to 4197 -- 4353\,\AA{}, 4373 -- 4440\,\AA{}, 5020 -- 5155\,\AA{}, 5476 -- 6165\,\AA{}, 6230 -- 6275\,\AA{}, and 6854 -- 7300\,\AA{}. We used a set of 53 high-quality stars from the Indo-US library as stellar templates \citep{valdes04, shetty15, guerou17}. This library provides high-enough resolution with regard to our spectral data, and fully covers the wavelength range of interest. We convolved the templates to an estimate of the average instrumental width and then used pPXF to fit the velocity dispersion for every spectrum. The pPXF method is particularly useful for correcting the emission line profiles for the underlying absorption features. In addition to the stellar templates, we added a power-law component $F_{\lambda} \propto \lambda^{-\alpha}$ to the fitting process in order to take into account the underlying non-stellar AGN continuum. For the exponent of the power law, we used a value of $\alpha = 0.76$. This value is based on the slope of the variable continuum in the rms spectrum (Fig.~\ref{IRAS23226_AVG_RMS_20221026_legend.pdf}).

\begin{figure*}
\centering
\includegraphics[width=1.0\linewidth,angle=0]{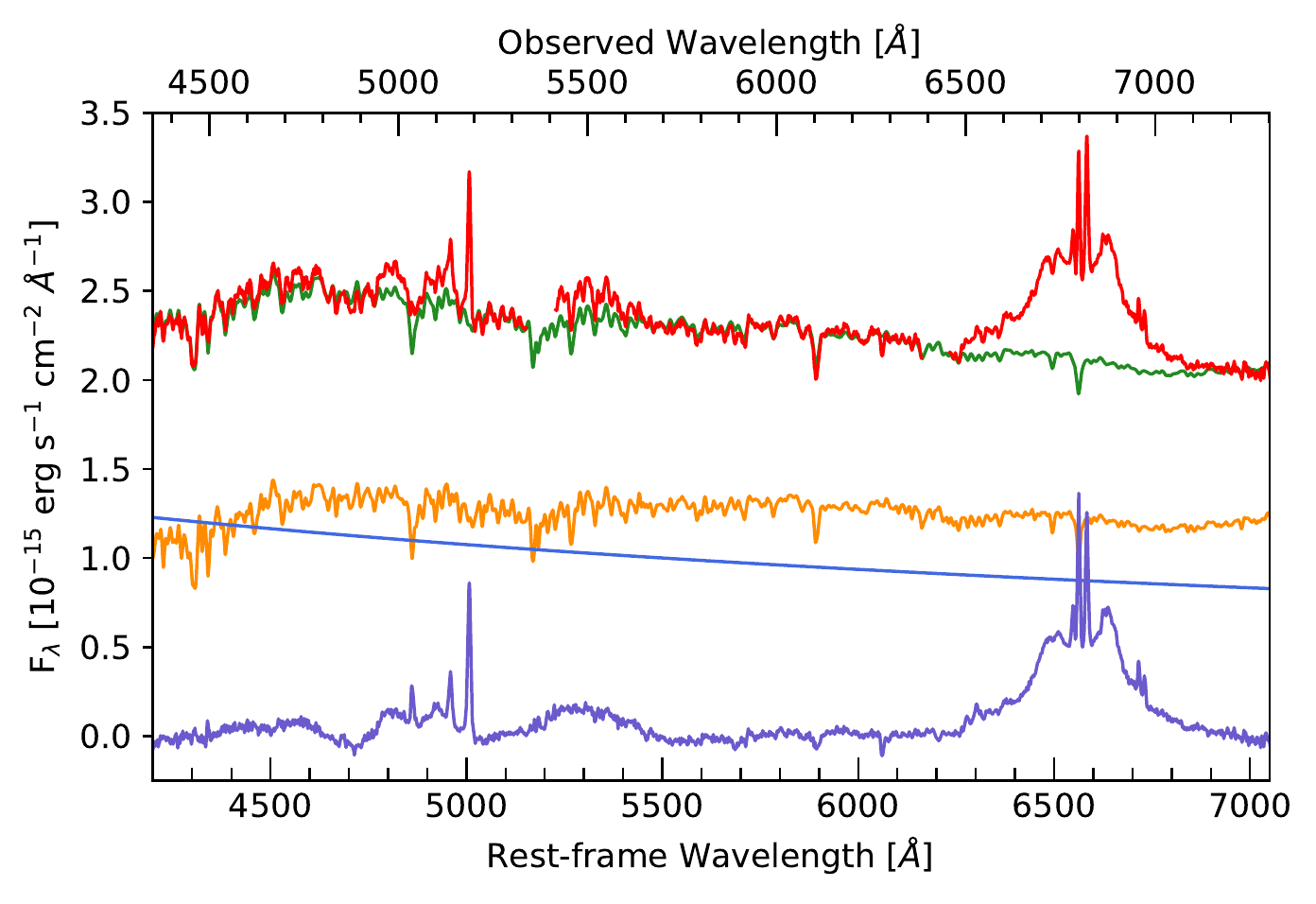}
\caption{Observed spectrum of \iras{} taken on 2019 September 24 (red) and synthesis fit with pPXF. The green line is the combined pPXF fit for the stellar component (orange) and the additional power-law component (light blue). The residuum spectrum (purple) i.e., the difference between the observed spectrum and the combined pPXF fit, gives the clean emission line contribution.}
\label{wkPPXF20190924_1.pdf}
\end{figure*}

We present the results of the spectral synthesis using pPXF for the observed spectrum of \iras{} taken on 2019 September 24 (red) in Fig.~\ref{wkPPXF20190924_1.pdf}. The green line is the combined pPXF fit for the stellar component (orange) and the additional power-law component (blue). The difference between the observed spectrum and the combined pPXF fit is the residuum spectrum  (purple)  that gives the clean emission line contribution.

\subsection{\bf Balmer emission-line profiles and their variations}

From the mean and rms spectrum presented in Fig.~\ref{IRAS23226_AVG_RMS_20221026_legend.pdf}, it is evident that the Balmer lines in \iras{} are very complex and broad. In particular, we see double-peaked profiles of \Ha{} and \Hb{} that are a signature of an accretion disk. In addition, \Ha{} shows  an underlying broad and blueshifted component. This very broad component is not identifiable in the \Hb{} line due to the superposition with the strong \ion{Fe}{ii} blends on the blue side of \Hb. Notably, \iras{} switched to an optical Seyfert 2 spectrum in 2020 showing no longer evidence for broad emission lines.

\begin{figure}
\centering
\includegraphics[width=1.0\linewidth,angle=0]{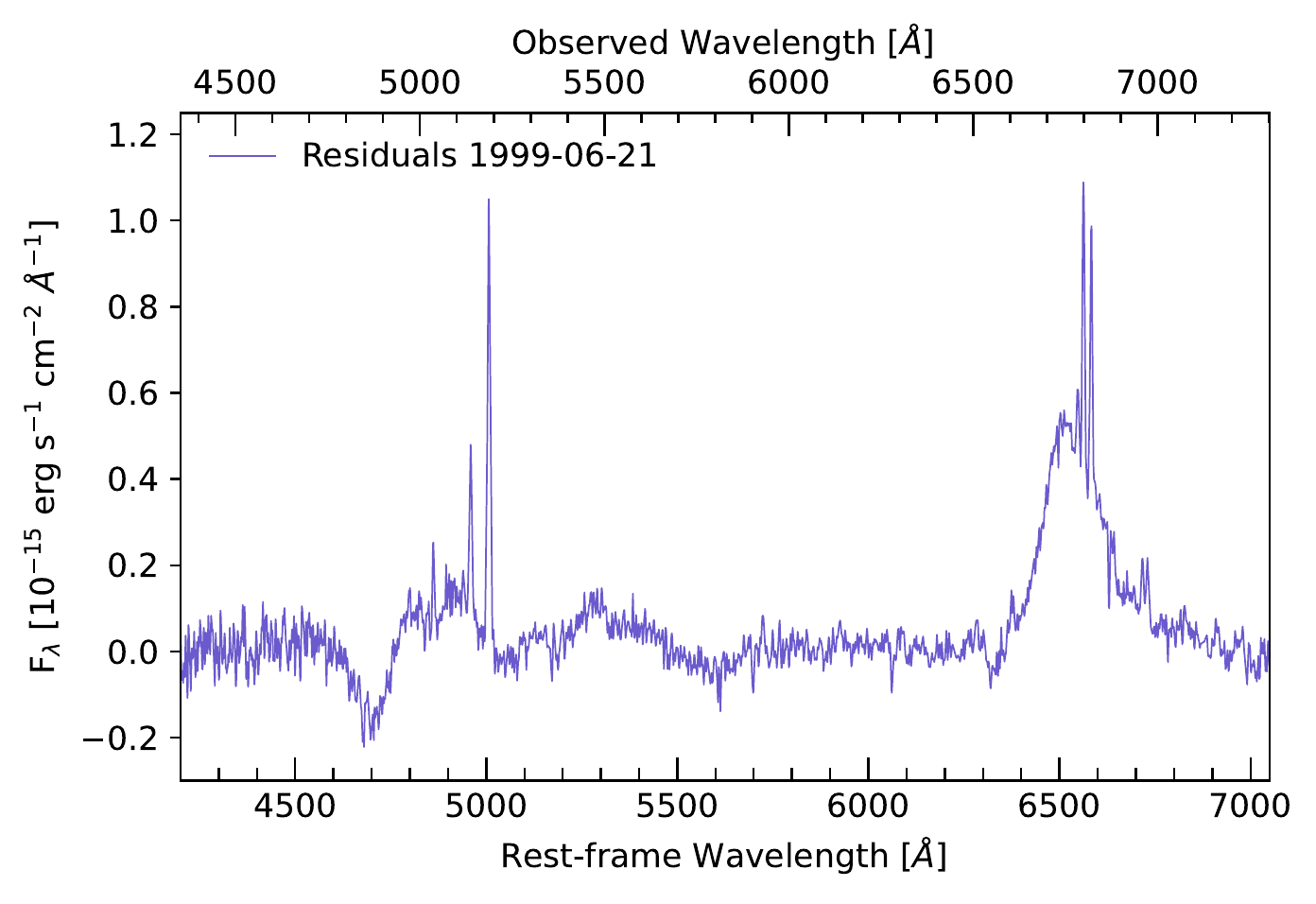}
\caption{Residuum spectrum of \iras{}  from 1991 June 21 indicating the Balmer absorption component in addition to the emission lines.}
\label{OchmIRAS23226_19990621_Grupe_20220325.pdf}
\end{figure}

There is an additional broad blue absorption component at $-10\,400\pm1000$\,\kms{} in \Ha{} and \Hb{} for all epochs except for the low states in 2017. This absorption was strongest in 1999. We show in Figure~\ref{OchmIRAS23226_19990621_Grupe_20220325.pdf} the emission-line residuum
spectrum of \iras{} from 1999 June 21 (purple), clearly indicating the additional Balmer
absorption components at $-10\,400$\,\kms.
The absorption seems to be stronger in \Hb{} than in \Ha{} due to the strong blue emission wing in \Ha{}. 

\begin{figure}
\centering
\includegraphics[width=1.0\linewidth,angle=0]{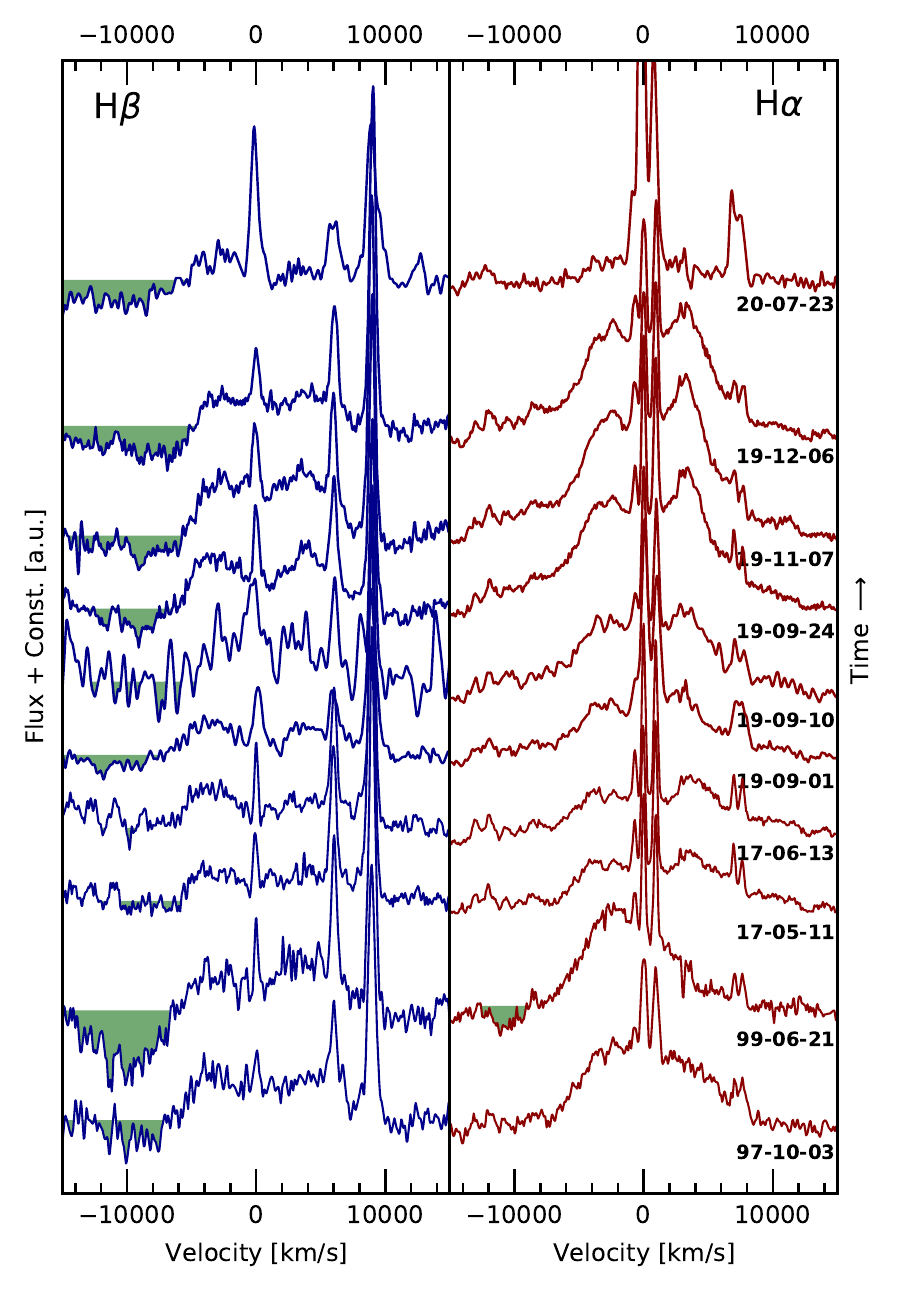}
\caption{Line profiles of \Ha{} and \Hb{} in velocity space
after subtraction of the host galaxy spectrum. Absorption components in the blue wing of the Balmer lines  (i.e., flux below zero) are shaded in green.}
\label{OchmIRAS23226_veloplots_20220325.pdf}
\end{figure}

A comparison of the \Ha{} and \Hb{} line profiles in velocity space is presented in Figure~\ref{OchmIRAS23226_veloplots_20220325.pdf}. As mentioned before, we first subtracted the spectrum of the host galaxy by means of a spectral synthesis using pPXF (see Sect.~\ref{sec:host_galaxy} and Fig.~\ref{wkPPXF20190924_1.pdf}) in order to obtain pure broad emission-line profiles.
The double peaks are present at all epochs before 2020 -- except for the red component of \Ha{} in the years 1997 and 1999. Profiles with double peaks can be produced through Keplerian rotation of accretion disks \citep{chen89a}, and we probe this interpretation later. The relative velocity of these
components remained constant over all the years. The blue peak originates at a velocity of $-3150\pm150$\,\kms{} and the red peak originates
at a velocity of $3200\pm150$\,\kms{} in the mean profiles of both lines.  

\begin{figure}
\centering
\includegraphics[width=1.0\linewidth,angle=0]{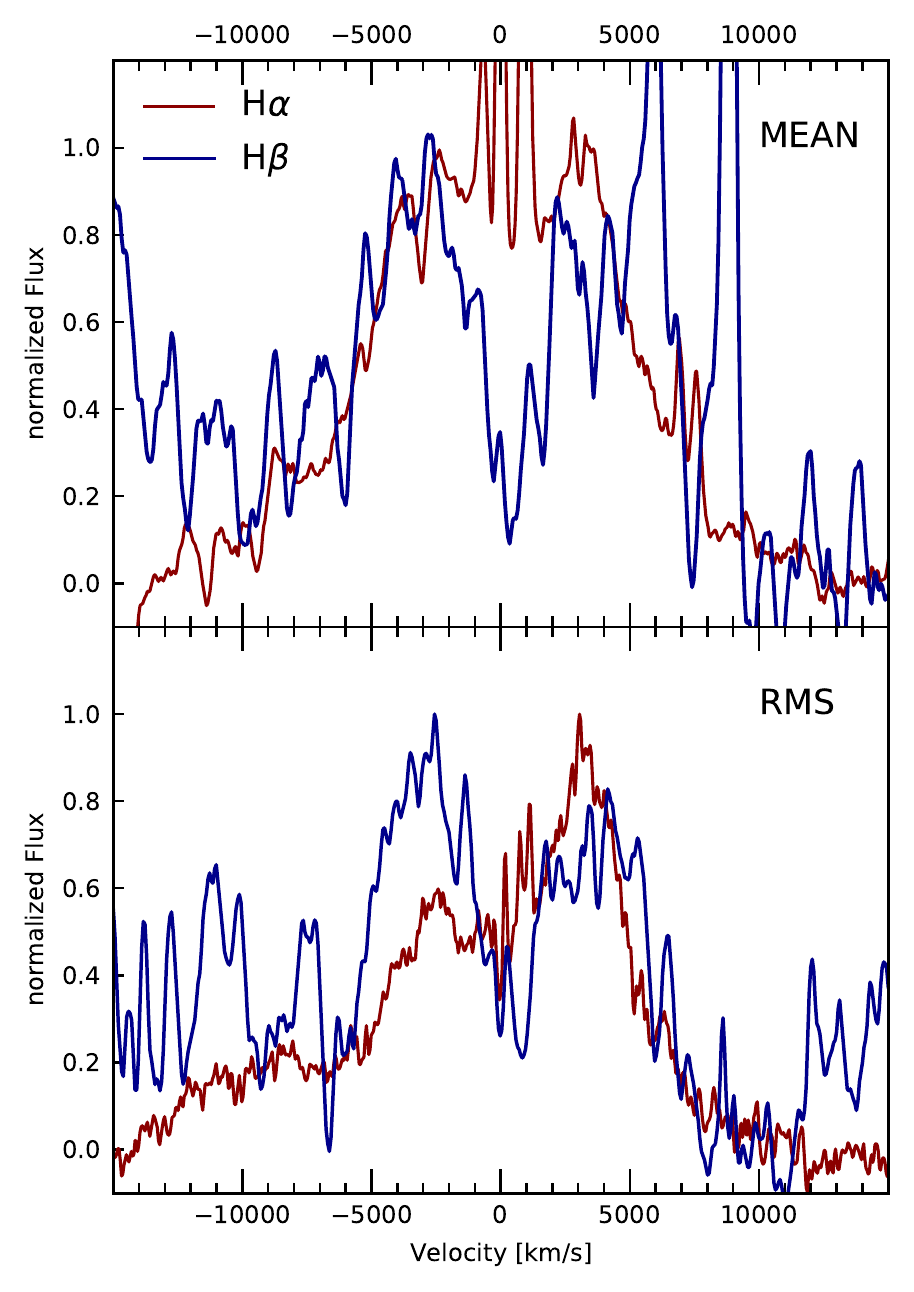}
\caption{Normalized mean and rms profiles of \Ha{} (red) and \Hb{} (blue) in velocity space.}
\label{OchmIRAS23226_velo_avgrms_20220329.pdf}
\end{figure}

We give the full width at half maximum  (FWHM) and the full width at zero intensity (FWZI) of the mean and rms profiles of \Ha{} and \Hb{} in Table~\ref{Balmerlinewidths}.  
\begin{table*}
{\small
\caption{Balmer line widths FWHM and FWZI, as well as the respective left wing boundary and right wing boundary position, and positions of the blue and red peaks of the mean and rms profiles.}
\label{Balmerlinewidths}
\begin{tabular}{l c c c c c c c c}
\hline \hline
Line & FWHM(left) & FWHM(right) & FWHM & FWZI(left) & FWZI(right) & FWZI & blue peak & red peak\\ 
\hline
\Ha{}$_{\rm mean}$\ & $-5470\pm200$ & $5340\pm200$ &  $10810\pm280$ & $-13570\pm400$ & $12200\pm400$ & $25770\pm570$ & $-3100\pm500$ & $3100\pm300$\\ 
H$\beta_{\rm mean}$\ & $-5680\pm300$ & $6500\pm1000$ &  $12180\pm1040$ & $<-6130\pm300$ & $7380\pm500$ & $>13510\pm580$ & $-3200\pm500$ & $3340\pm500$\\
\hline
\Ha{}$_{\rm rms}$\ & $-3190\pm500$ & $4860\pm300$ &  $8050\pm580$ & $-14120\pm400$ & $11770\pm500$ & $25890\pm640$ & $-2500\pm300$ & $3200\pm300$\\ 
H$\beta_{\rm rms}$\ & $-5170\pm300$ & $5640\pm400$ &  $10810\pm500$ & $<-6630\pm400$ & $7800\pm1000$ & $>15630\pm1080$ & $-2900\pm400$ & $3520\pm400$\\
\hline
\end{tabular}}
\tablefoot{In units of {\kms}.}

\end{table*}
The FWHM of \Hb$_{\rm mean}$ and \Ha$_{\rm mean}$ amounts to $\sim 11\,000 \pm 1000$\,\kms. The FWZI of \Ha{} amounts to $\sim 26\,000$\,\kms. We can only give a lower limit of $13\,500$\,\kms{} for the FWZI of \Hb{} due to the strong blue absorption and the blending of the blue wing with the \ion{Fe}{ii} blend.
 
A comparison of the individual \Ha{} and \Hb{} line profiles shows differences in the profiles for different epochs. The \Ha{} profile was asymmetric with respect to $v=0$\,\kms{} for the early epochs in 1997 and 1999. The red component was much weaker than the blue component at these epochs. In contrast, \Ha{} showed almost symmetrical double-peaked profiles for the later epochs in 2017 and 2019 independent of the integrated line flux. The \Hb{} profile showed double-peaked profiles at all epochs.

\subsection{\bf X-ray spectrum }

\begin{figure}
    \centering
    \includegraphics[width=\linewidth]{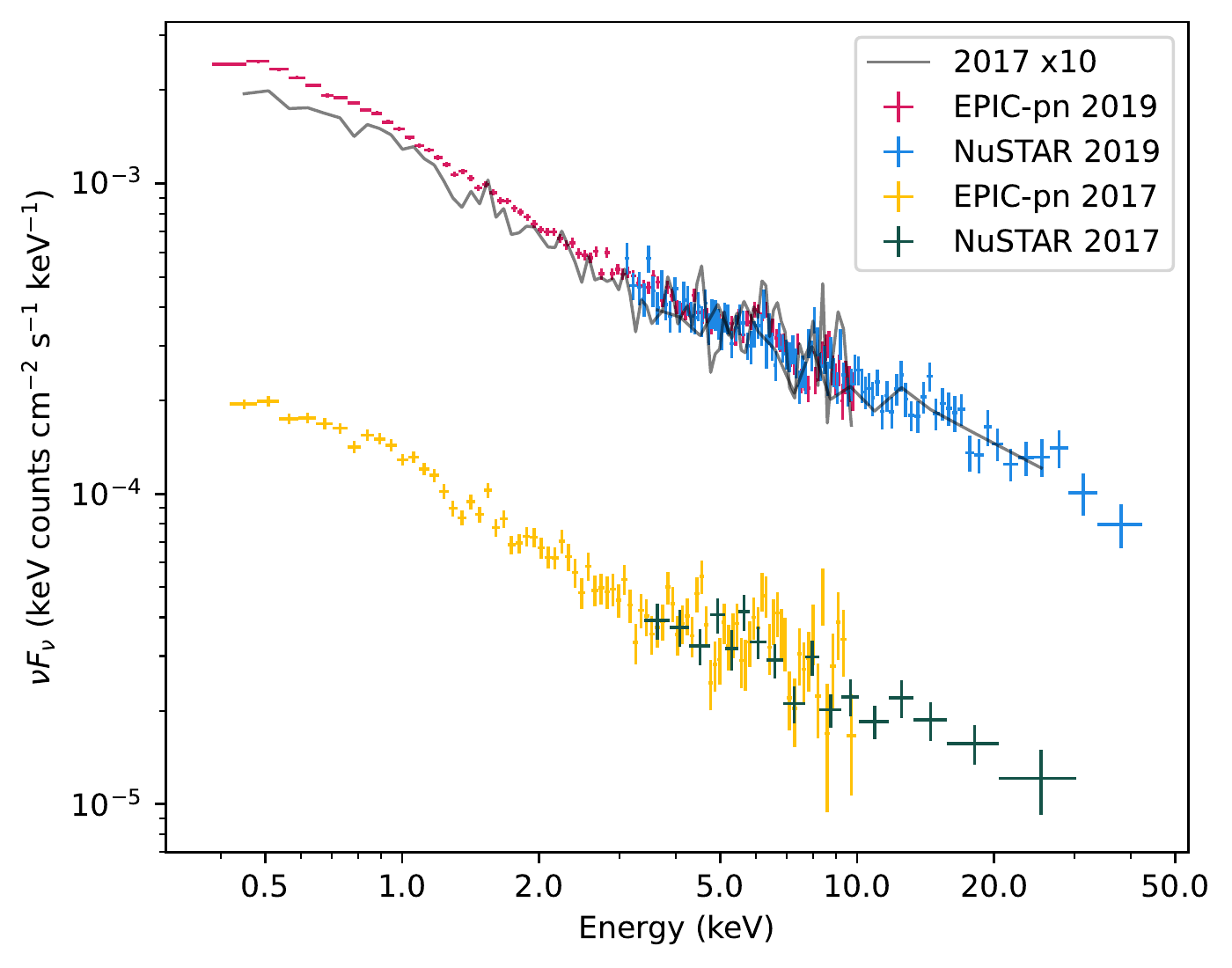}
    \caption{Comparison of the broadband X-ray data in 2017 and 2019. The change in flux is a uniform increase of a factor of $\sim10$ across the bandpass, and both spectra show residuals in the Fe~K band. All spectra are corrected for the instrumental effective area. The gray line shows the 2017 data scaled up by a factor of 10 to match the 2019 data. Above $\sim3$~keV the agreement in spectral shape is very good, below 3~keV the 2019 data is relatively higher, with the difference increasing toward lower energies.}
    \label{fig:xray_comparison}
\end{figure}

\begin{figure}
    \centering
    \includegraphics[width=\linewidth]{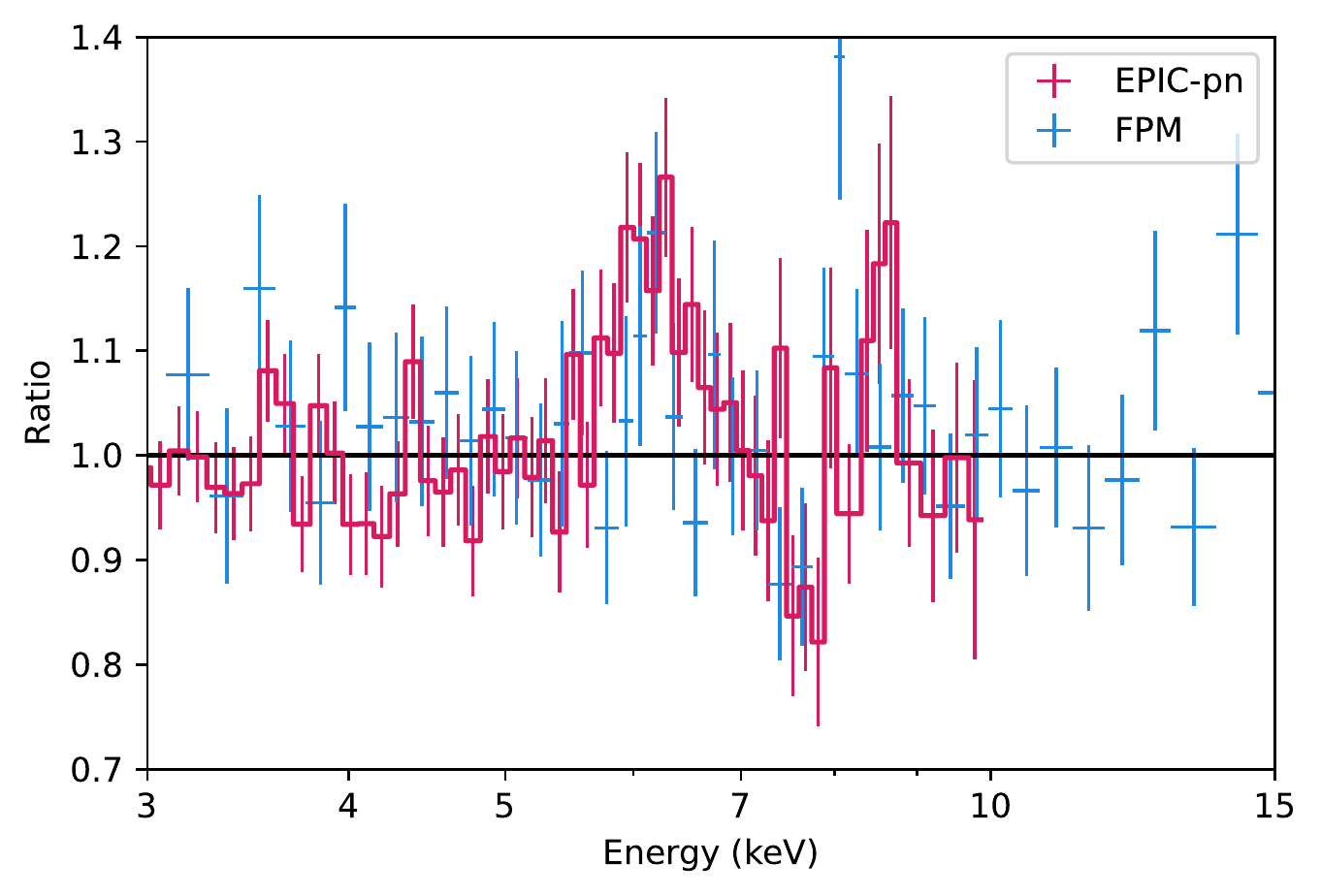}
    \caption{Ratio of the \xmm\ and \nustar\ spectra to a power-law continuum and Comptonized soft excess model, in the Fe~K band. A moderately broadened emission line is visible, and a possible absorption feature is present between 7 and 8 keV in the EPIC-pn spectrum. The NuSTAR data are consistent with both features, but are noisier.}
    \label{fig:xray_fe_ratio}
\end{figure}

The broadband X-ray spectrum is qualitatively similar to that in 2017, but a factor of 10 brighter across the whole bandpass (see Fig.~\ref{fig:xray_comparison}). A weak, smooth soft excess is present, along with a weak Fe K emission feature and a possible absorption feature in the Fe band (see Fig.~\ref{fig:xray_fe_ratio}) at a slightly higher energy (7.6~keV) than the possible feature seen in the 2017 data ($\sim7.3$~keV). If genuine, this would correspond to an outflow with velocity of $\sim0.12c$ (36000~km~s$^{-1}$). Highly ionized absorbers with outflow velocities higher than 10\,000~km~s$^{-1}$ have been defined as ultra-fast outflows \citep[UFOs;][]{tombesi10}. The soft X-ray band appears featureless, so we first test modeling it with a simple Comptonization model with no atomic features \citep[e.g.,][]{Petrucci18}. We find that the soft excess is well modeled with such a Comptonization model. A broadband fit with a power-law continuum, Comptonized soft excess \citep[\textsc{nthcomp}][]{Zdziarski96,Zycki99}, and Galactic absorption \citep[\textsc{tbabs};][]{Wilms00} gives a good fit to the combined EPIC-pn and NuSTAR spectrum ($\chi^2$/dof of $343/286=1.20$, or $219/208=1.05$ if we exclude the 5--8~keV band). While other models have been proposed for the soft excess, most commonly reflection \citep[e.g.,][]{Jiang19}, the lack of any structure in the soft excess means that more complex models will either be disfavored or unconstrained with the data available, so we assume a Comptonized origin.

While this simple model explains the continuum shape very well, it does not account for the high energy complexity. We therefore focus on modeling the residuals in the Fe~K band. We consider two main possibilities for the moderately broadened iron line, where the emission either is produced in reflection from the accretion disk \citep[e.g.,][]{Fabian89} or scattered from a wind \citep[e.g.,][]{Nardini15}. It is worth noting that the emission from these two processes is highly degenerate \citep[][]{Parker22}, and it is entirely possible that some contribution from both is present in the spectrum. For this reason, we also consider a hybrid model, where X-rays are first reflected from the disk and then pass through a wind which produces further Fe~K emission. We use \textsc{tbabs} \citep{Wilms00} with a column density of $1.58\times10^{20}$~cm$^{-2}$ \citep{HI4PI16} to model Galactic absorption, \textsc{nthcomp} \citep[][]{Zdziarski96,Zycki99} for a Comptonized soft excess, \textsc{relxill} and \textsc{xillver} \citep{Garcia13, Garcia14} for relativistic and distant reflection, and a table of disk wind spectra based on the simulations of \citet{Sim08,Sim10} \citep[this is the same version of the model used in][]{Parker22}.
In \textsc{xspec} format, the models are \textsc{tbabs $\times$ (xillver + gabs $\times$ (nthcomp + powerlaw + relxill))} for the reflection model, \textsc{tbabs $\times$ (xillver + diskwind $\times$ (nthcomp + powerlaw))} for the disk wind model, and \textsc{tbabs $\times$ (xillver + diskwind $\times$ (nthcomp + powerlaw + relxill))} for the hybrid model.
All three models give an acceptable fit to the data ($\chi^2$/dof of 287/277, 306/281, and 276/275 for reflection, disk wind, and hybrid models respectively), without leaving obvious residual structures.

\begin{figure*}
    \centering
    \includegraphics[width=0.45\linewidth]{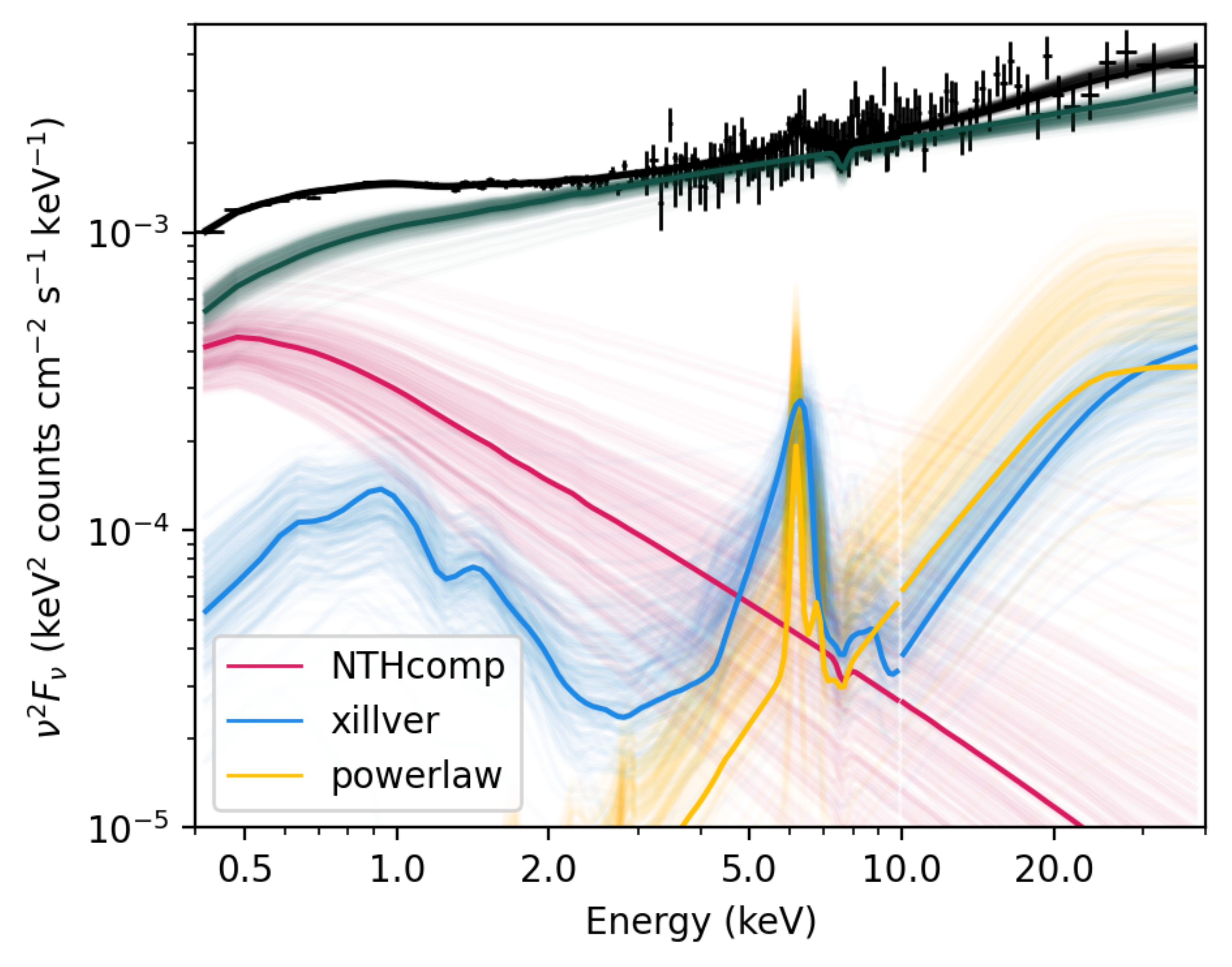}
    \includegraphics[width=0.45\linewidth]{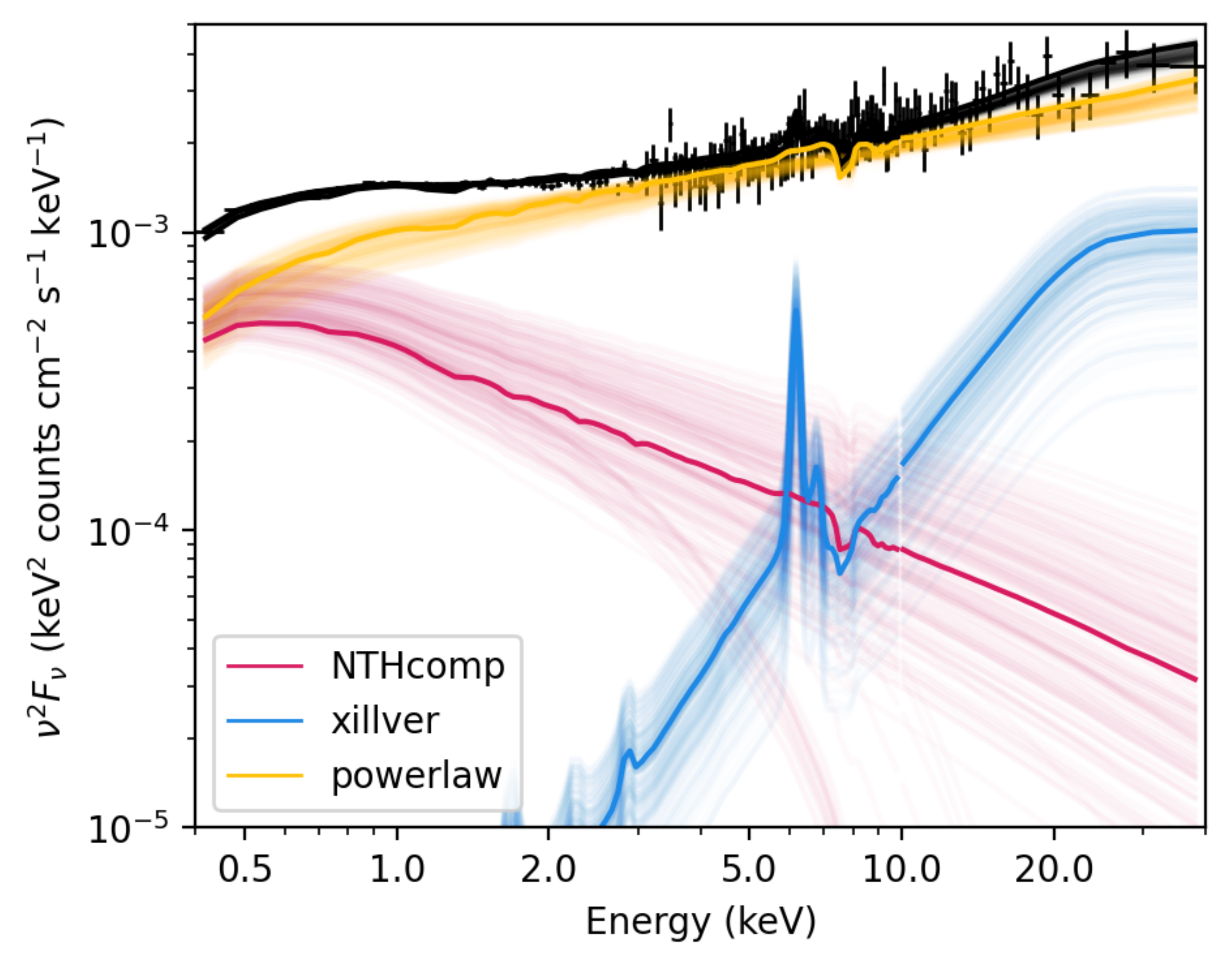}
    \includegraphics[width=0.45\linewidth]{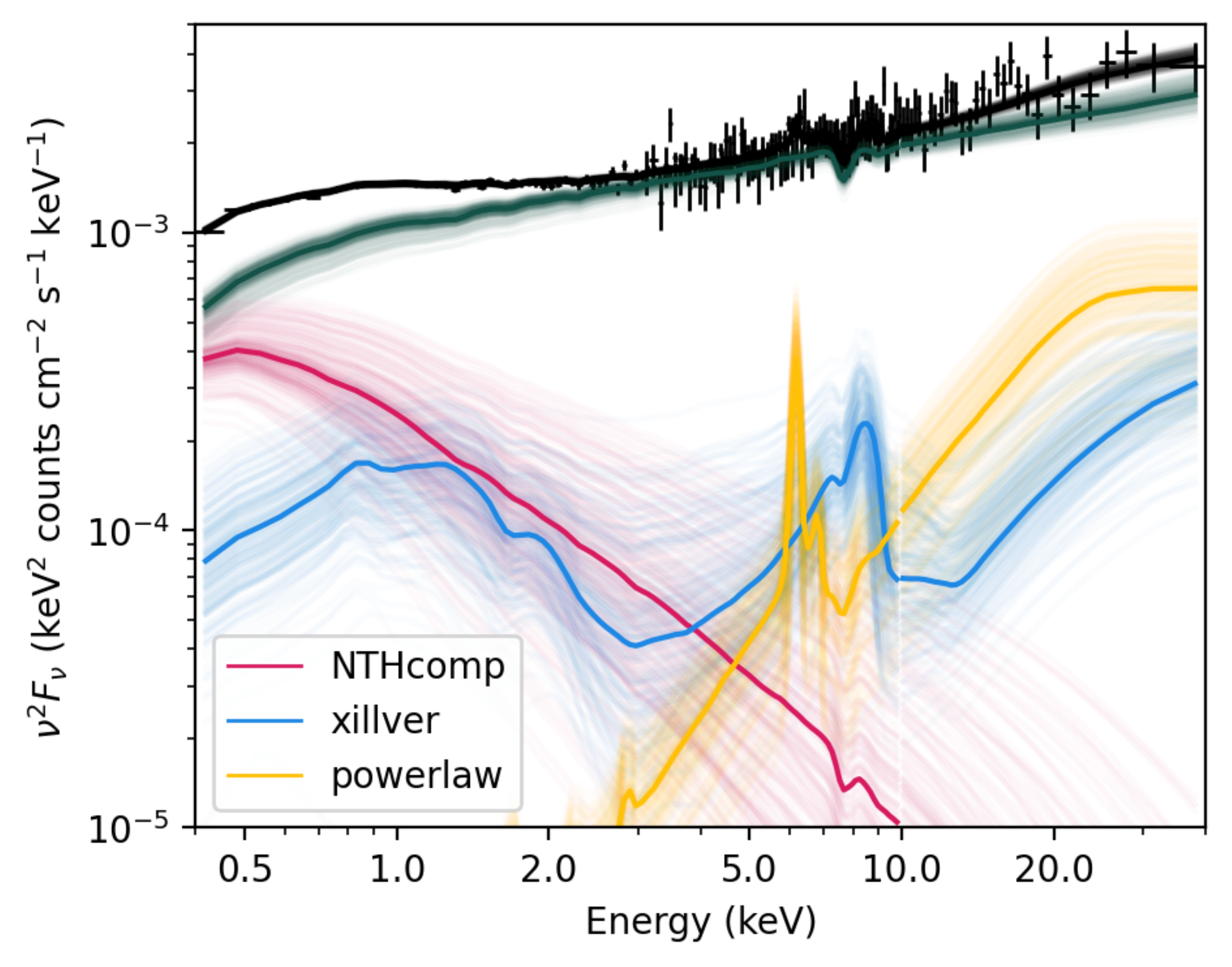}
    \caption{Model spectra for the three different X-ray models, fit to the 2019 \xmm\ EPIC-pn spectrum. Top left shows the reflection model, top right the disk wind model, and bottom the hybrid model. In each case, we plot 200 model spectra with parameters randomly selected from the MCMC chains as transparent lines, and the model corresponding to the peak of the posterior distribution for each parameter as solid lines. This effectively plots the posterior distribution of models, showing the uncertainty in each model component. In each case, the points with error bars show the data, and the black lines show the total model spectra. For the disk wind and hybrid models, the disk wind component is included as a multiplicative component applied to all other spectra components apart from the distant reflection. In the reflection model we apply a Gaussian absorption line to all additive components aside from the distant reflection to account for the possible absorption feature.}
    \label{fig:xray_models}
\end{figure*}

Having established baseline fits, we run MCMC on each fit to explore the parameter space. We use a chain length of $5\times10^5$, 200 walkers, and a burn in period of $5\times10^4$ steps. We then calculate the most likely parameter values and one-sigma errors from the posterior distributions of the chains (Table~\ref{tab:xray_pars}). In Fig.~\ref{fig:xray_models} we show the different models used to fit the data, in each case we plot a random sample of 200 model spectra drawn from the MCMC chains, showing the uncertainty in spectral shape of the different model components.

\begin{table*}
    \centering
    \caption{X-ray model parameters.}
    \begin{tabular}{l l c c c r}
    \hline
    \hline
         Model & Parameter & Reflection & Disk wind & Hybrid & Description (unit) \\
    \hline
         \textsc{tbabs} & $N_\mathrm{H}$ & $1.58\times10^{20}$ & $1.58\times10^{20}$ & $1.58\times10^{20}$ & Galactic column (cm$^{-2}$) \\
    % \hline
    \\
         \textsc{xillver} & $N$ & $(4.1_{-0.4}^{+5.2})\times10^-6$ &$(1.5_{-0.5}^{+0.4})\times10^{-5}$ & $(1.2_{-0.4}^{+0.5})\times10^{-5}$ & normalization$^2$\\
    % \hline
    \\
         \textsc{gabs} & $E$ & $7.6\pm0.1$ & - & - & line energy (keV)\\
                       & $N$ & $0.05_{-0.02}^{+0.03}$ & - & - & line depth\\
    % \hline
    \\
         \textsc{nthcomp} & $\Gamma$ & $3.0_{-0.2}^{+0.1}$ & $2.7\pm0.1$ &  $3.1_{ - 0.3}^{ + 0.2}$ & photon index\\
                          & $kT_\mathrm{e}$ & $134_{-110}^{+244}$ &$>95$ & $4_{-2}^{+4}$ & electron temperature (keV)\\
                          & $N$ & $(3\pm1)\times10^{-4}$ &  $(5.8_{-0.6}^{+2.7})\times10^{-4}$ & $(4_{-1}^{+2})\times10^{-4}$ & normalization \\
    % \hline
    \\
         \textsc{relxill} & $q$ & $<4.5$ & - & $<5$ & emissivity index\\
                          & $a$ & $<-0.45$ & - & $0.5_{-0.5}^{+0.3}$ & spin\\
                          & $i$ & $25_{-2}^{+13}$ & - & $80_{-3}^{+2}$ & inclination (degrees)\\
                          & $\log(\xi)$ & $2.97_{- 0.11}^{+0.05}$ & - & $2.86_{- 0.08}^{+ 0.17}$ & ionization (erg~cm~s$^{-1}$)\\
                          & $A_\mathrm{Fe}$ & $>8.4$ & - & $8.5_{-2.4}^{+1.4}$ & iron abundance (solar)\\
                          & $N$ & ${(6\pm1)}\times10^{-6}$ & - & ${(2.6_{-0.6}^{+0.9})}\times10^{-6}$ & normalization$^2$ \\
    \\
         \textsc{diskwind} & $\dot{M}$ & - &  $0.05_{ - 0.03}^{ + 0.12}$ & $0.039_{-0.006}^{+0.012}$ & mass outflow rate ($\dot{M}_\mathrm{Edd}$)\\
                           & $f_v$ & - & $1.25^1$ & 2.0$^1$ & terminal velocity parameter\\
                           & $L_\mathrm{X}/L_\mathrm{Edd}$ & - & $< 0.11$& $<0.04$ & X-ray luminosity\\
                           & $i$ & - & $62^1$ & $74^1$ & inclination (degrees)\\
    % \hline
    \\
         \textsc{powerlaw} & $\Gamma$ & $1.74_{ - 0.03}^{ + 0.04}$ &  $1.71\pm0.05$ & $1.76_{-0.04}^{+0.03}$ & photon index\\
                           & $N$ & $(11\pm1)\times10^{-4}$ & $(14_{-1}^{+2})\times10^{-4}$ & $(17\pm1)\times10^{-4}$ & norm. (keV$^{-1}$~cm$^{-2}$~s$^{-1}$) \\
    \\
         \textsc{constant} & $C_\mathrm{FPM}$ & $1.08\pm0.02$ & $1.07_{-0.01}^{+0.02}$ & $1.07\pm0.02$ & FPM/EPIC-pn ratio\\
    \\
         & $\chi^2$/dof & $287/277=1.03$ & $306/281=1.09$ & $276/275=1.00$ & fit statistic\\
    \hline
        
    \end{tabular}
    \label{tab:xray_pars}
\begin{flushleft}    
$^1$In this version of the disk wind model the $f_v$ and $\mu=cos(i)$ parameters do not interpolate well between grid points, so we fix these parameters to the closest grid point after the intial fit, and do not calculate errors. This issue is fixed in the new version of the model \citep{matzeu22}.

$^2$The normalizations of the \textsc{xillver} and \textsc{relxill models} are defined based on the flux of the incident spectrum, see \citet{Dauser16}.
\end{flushleft}
\end{table*}

We can estimate the significance of the possible absorption line by examining the change in fit statistic when including or excluding the Gaussian line in the reflection model (the disk wind model describes the absorption line, but also the emission and other features throughout the band, so cannot test for a single feature). Removing the Gaussian worsens the fit by $\Delta\chi^2=5$, for 2 fewer degrees of freedom in the model. This is not highly significant ($\sim2\sigma$), even without taking into account the look-elsewhere effect. Based on these results, we can conclude that some Fe emission from either a mildly relativistic wind or the inner accretion disk is required to explain the observed data, but we cannot conclusively determine which process dominates.

\section{Discussion}

\subsection{\bf X-ray, UV, and optical continuum variations based on \swift}

The X-ray, UV, and optical continuum observations of \iras{} with \swift{} all show the same general variability trend from 2007 to 2021 (see Figs.~\ref{fig_dirk_xrt_uvot_lc07_19.pdf} and \ref{fig_dirk_xrt_uvot_lc19.pdf}). The variability behavior can be summarized as follows:
We observe a general slow flux decrease by a factor of 5 in all bands from 2007 until July 2016.
Subsequently, \iras{} showed a decrease in X-ray flux by a factor of about 4
from July 2016 until July 2017.
Afterwards, the X-ray flux increased to even higher
intensity level than before by a factor of $\sim$ 30 until August 2019.
After a short decline, \iras{} showed an outburst in the continuum lasting from 2019 September  until 2019 December  (see Fig.~\ref{fig_dirk_xrt_uvot_lc19.pdf}). This outburst was observed spectroscopically by accompanying optical observations with the SAAO\,1.9\,m telescope and SALT (see Fig.~\ref{Ochmlightcurves_IRAS23226_20220329.pdf}) in 2019.
The overall flux increased again in December 2019 and finally decreased in 2021.

The decreasing and increasing variability trends were strongest in  X-rays and less pronounced in the optical B and V bands. However, one has to consider that the U, B, and V \swift{} bands were not corrected for the strong contribution of the host galaxy. In order to quantify the variability strength, we calculated the fractional variation $F_{\rm var}$ for each \swift \ band in Sect.~\ref{sec:swift_results}.
The results for the X-ray, UV, and optical bands are presented in Tab.~\ref{swiftvarstatistics}. The fractional variations deduced from the \swift{} observations from 2007 until 2021 follow the same trend, that is to say decrease in fractional variation with increasing wavelength, as those based on the shorter period from 2007 until 2017 only \citep{kollatschny20}. The fractional variations in the UV and optical bands in \iras{} are stronger than those in, for example, the changing-look AGN HE\,1136-2304 \citep{zetzl18}, and the prototype AGN NGC\,5548 \citep{edelson15, fausnaugh16}. More precisely, the fractional variations in \iras{} are stronger by a factor of 2 in comparison to the changing-look AGN HE\,1136-2304, and even more so with regard to NGC\,5548.

\subsection{\bf X-ray spectrum}

%- Comparison of the deep broadband \xmm/\nustar\ spectra taken during the minimum in 2017 and at the maximum in 2019.\\ 
The 2019 broadband X-ray spectra of \iras{} are qualitatively very similar to the 2017 spectrum, but a factor of 10 brighter across the bandpass, with no large change in spectra hardness and no dramatic evolution of the soft excess sometimes seen in changing look AGN \citep[e.g.,][]{noda18,parker19}. The RMS spectrum is consistent with being flat, with no significant absorption or emission features visible.

%- respective relative contribution of the different X-ray components.\\
In the previous paper we modeled the 2017 data with distant reflection, a Comptonized continuum, a warm absorber, and a phenomenological black body soft excess. The 2019 data is of higher quality due to the higher flux, and enabled us to rule out the presence of a warm absorber based on the featureless RGS spectrum. While it is possible that this component was present in 2017 and not in 2019, we consider it more likely that no warm absorber has been present throughout and this component was fitting some other aspect of the spectrum in 2017, such as the spectral complexity introduced by a more complex soft excess including Comptonization and reflection.

%- relativistic outflow at $\sim7.3$~keV at both epochs?\\
We have found low significance ($\sim2\sigma$ in each case) evidence for an absorption line from an outflow in both 2017 and 2019 spectra, at 7.3 and 7.6~keV (observed frame). 
We see evidence for a complex iron emission line profile, broadened beyond the level that could be produced by distant reflection alone. We consider relativistic reflection and disk wind models for this emission, along with a combination of both. It is not obvious which model should be preferred: the hybrid model gives the best fit, but both wind and reflection components prefer high inclinations (74 and 80 degrees, respectively) which is hard to reconcile with the unabsorbed spectrum. The reflection model gives a lower inclination (25 degrees), which is consistent with the lack of absorption, but also requires retrograde spin and a highly super-solar iron abundance ($>8.4$), although we note that this may be due to the fixed disk density assumed, as high density reflection can result in lower iron abundances \citep[e.g.,][]{Tomsick18,Jiang19}. The pure disk wind model parameters are more reasonable, with an inclination of 62 degrees and no requirement for extreme parameters, but it also gives a worse fit to the data. This may simply be due to the more limited parameter space of this model relative to the reflection model.

\subsection{\bf Optical spectrum: Extremely broad and double-peaked Balmer profiles}
 
\iras{} was originally classified as a Seyfert 1 galaxy  \citep{allen91} based on an optical spectrum taken between 1985 and 1990. It was of clear Seyfert type 1 in the years 1997 and 1999 (see Figs.~\ref{IRAS23226_all_spectra_20220326.pdf} and  \ref{OchmIRAS23226_veloplots_20220325.pdf}), and the broad \Ha{} line profile was asymmetric at both observing epochs, showing 
a strong blue peak originating at a velocity of $-3100\pm500$\,\kms{} without a corresponding red peak. Later on, however,
\Ha{} showed almost symmetrical double-peaked profiles for all observing epochs in 2017 and 2019, independent of the integrated line flux. The red peak originated at a velocity of $3100\pm300$\,\kms{}, hence at the mirrored velocity of the blue peak. In contrast to \Ha{}, the \Hb{} profile shows a double-peaked profile for all epochs between 1997 and 2020.  Taking into account the measurement uncertainties, the positions of the blue and red peaks in velocity space were identical in both \Ha{} and \Hb{}, and they remained constant in all our spectra. In 2020, the spectral type of \iras{} changed to that of a Seyfert 2 type 
(see Fig.~\ref{OchmIRAS23226_veloplots_20220325.pdf}).

During our campaign, the line profiles in the red and blue wings of \Ha{} and \Hb{} varied with different amplitudes. This can be seen best in the rms line profiles (see Figs.~\ref{IRAS23226_AVG_RMS_20221026_legend.pdf},  \ref{OchmIRAS23226_veloplots_20220325.pdf}, and \ref{OchmIRAS23226_velo_avgrms_20220329.pdf}). \Ha{} shows considerably stronger variations in the red wing than in the blue wing from 1997 until 2017. Afterwards, the blue and red wing varied uniformly.

The broad Balmer line components were faint and barely visible in 2017 in comparison to the optical spectra taken in 1997, 1999, as well as in 2019 (see Fig.~\ref{OchmIRAS23226_veloplots_20220325.pdf}). Especially, the broad \Hb{} component
was barely visible without correction for the host galaxy. It is usually presumed that broad double-peaked emission lines are generated in relativistic Keplerian disks of gas surrounding the central supermassive black hole, as models developed for this physical scenario produce such profiles \citep[e.g.,][]{chen89a,chen89b}. During the minimum state in 2017 as well as during the maximum state in 2019, the two Balmer lines (\Ha{} and \Hb) show the same symmetric double-peaked profile. The peak separation corresponds to 6200 \kms{}. These values are consistent with the peak velocities of many other AGN emission line profiles previously fitted with Keplerian disk models \citep{eracleous94,eracleous03}. In cases with such peak separation values, these earlier studies achieved optimal fits by choosing an accretion disk inner radius of $\sim 500$ Schwarzschild radii.

In our present study, we derive higher Balmer decrement values in comparison to the low values in our earlier investigation in \cite{kollatschny18}. This is due to the \Hb{} emission-line flux  being weaker now after correction for the modeled stellar host component. Furthermore, we now take the additional  absorption component on the blue side of \Hb{} into account. The Case B recombination  value of the Balmer decrement \Ha{}/\Hb{} is 2.7  \citep{osterbrock06}. Here we derive Balmer decrement values of 3.7 up to 9, indicating dust extinction or optical
depth effects in the broad-line region.
The values for the Balmer decrement in \iras{} are similar to those in other AGN: \cite{antonucci83} found an anticorrelation of the Balmer decrement with continuum flux in NGC\,4151.
We observed that the Balmer decrement \Ha/\Hb{} increases with the decreasing \Hb{} line flux in variable Seyfert galaxies, such as Mrk\,110 (\citealt{bischoff99}), NGC\,7603 (\citealt{kollatschny00}), and HE\,1136-2304 (\citealt{kollatschny18}). This general trend is also present in \iras{}. However, the associated errors are larger in \iras{} than in those galaxies. 
There is evidence that the radial structure \citep[e.g.,][]{koratkar91,kaspi05} and the vertical structure \citep{kollatschny14} of the line emitting BLR clouds vary as a function of continuum luminosity. This can lead to variations in the Balmer decrement.
A compilation of Balmer decrement data is given \cite{jaffarian20}. A slightly higher \Ha{}/\Hb{} value in AGN with double-peaked Balmer lines can also be a consequence of higher reddening in these AGN which are assumed to be seen at higher inclination (\citealt{gaskell17}).

The asymmetry of \Ha{} during the early states in 1997 and 1999 might have been caused by absorption of the red component.  Partial dust obscuration of the BLR by outflowing dust clumps, a model introduced by \citet{gaskell18}, can produce asymmetries and velocity-dependent lags. Other models to explain variable profile humps are conical outflows \citep{zheng91}, an asymmetric distribution of the BLR clouds \citep{wanders96}, and/or asymmetric disks \citep{eracleous94}.

There are clear signs for a blue absorption component in \Hb{} and \Ha{} at $-10\,400\pm1000$,\kms{} (see Figures~\ref{OchmIRAS23226_19990621_Grupe_20220325.pdf} and
\ref{OchmIRAS23226_veloplots_20220325.pdf}). Furthermore, based on the X-ray spectra, there is additional evidence for an outflow component. This absorption feature is present in the Fe band (see Fig.~\ref{fig:xray_fe_ratio}) at a slightly higher energy (7.6~keV) 
than the possible feature seen in the 2017 data ($\sim7.3$~keV). It corresponds to an outflow with velocity of $\sim0.12c$ ($-36\,000$\,\kms{}).

The broad double-peaked Balmer line profiles and their strong variations in \iras{} (see Fig.~\ref{OchmIRAS23226_veloplots_20220325.pdf}) are 
similar to those of other broad double-peaked galaxies such as NGC\,1097 \citep{storchi97}, and NGC\,7213 \citep{schimoia17}, or broad-line radio galaxies such as 3C390.3 \citep{shapovalova10} and Arp102B \citep{shapovalova13}. \cite{strateva03} investigated the properties of double-peaked Balmer lines in a large sample of AGN based on SDSS galaxies. However, in comparison to all these double-peaked AGN, \iras{} shows significantly stronger \ion{Fe}{ii} line blends, in contrast to what is expected from Eigenvector 1 studies of AGN \citep{boroson92, sulentic00, shen14}.

\subsection{\bf Variability pattern in \iras{} }

The changing-look character of \iras{} is deduced from strong long-term (many years) and short-term (a few weeks) continuum variations in the optical and X-ray. In addition, \iras{} showed strong line profile variations including a transition from a Seyfert 1 type spectrum to a Seyfert 2 type in 2020.

\iras{} underwent a strong outburst lasting from 2019 September until 2019 December. The optical
continuum flux -- based on SALT spectra -- increased by a factor of 1.6 within two months. This corresponds to a factor of 3 when correcting the continuum flux for the host galaxy contribution 
(1.1 $\times$ 10$^{-15}$\,erg\,s$^{-1}$\,cm$^{-2}$\,\AA$^{-1}$ in the V band). \iras{} varied 
in the X-rays (\swift{}) by a factor of nearly 5 during these three months. The optical line intensities of  \Ha, \Hb, and the \ion{Fe}{ii}(42,48,49) blend varied by factors of 2.3, 2.4, and 4. However, the shape of the double-peaked line profiles of \Ha{} and \Hb{} remained more or less unchanged despite the strong intensity variations during these three months.

We took deep broadband \xmm/\nustar\ spectra during the optical maximum state (2019 November). These spectra can be compared with those taken during the minimum state in 2017 (see Fig.~\ref{Ochmlightcurves_IRAS23226_20220329.pdf} for 
the observing epochs). The 2019 X-ray spectra of \iras{} are qualitatively very similar to
the 2017 spectra, but brighter by a factor of 10. The X-ray spectra were unabsorbed in 2019 as well as
during the minimum state in 2017. The optical line profiles at these two epochs were also qualitatively identical despite intensity variations by a factor of more than two (see Fig.~\ref{OchmIRAS23226_veloplots_20220325.pdf}).

The optical line profiles, however, changed on timescales of several years. These line-profile changes are independent of the continuum flux intensities. In total, we can discern three separate time intervals, namely the years 1997 - 1999, 2017 - 2019, and 2020. The line profiles of \Ha{} showed asymmetric blue excess profiles during the years 1997 and 1999. However, the profiles  were symmetric and double peaked during the years 2017 and 2019. Finally, \iras{} changed to a Seyfert 2 type in 2020. A similar behavior has been observed in
the changing-look AGN HE\,1136-2304 \citep{zetzl18}: The line profiles varied independently of the continuum variations on timescales of years. The strong outburst observed in \iras{} in autumn 2019 is embedded
in additional variations before and after that epoch.

The cause for the violent as well as long-term variations in  \iras{} can be narrowed down based on the observed long-term and short-term light curves in the X-ray and in the optical, the X-ray spectral variations, as well as optical line intensity and line profile variations.  
Microlensing or a tidal disruption event (TDE) can be excluded as an explanation for the long-term variability pattern in \iras. First, lensing would be monochromatic, while the multiband light curves differ in detail in different bands, the flaring repeats, and lensing would not change the Balmer-line profiles. Second, TDEs in gas-rich environments or illuminated streams of stellar debris sometimes do in fact show a temporary emission-line signal in form of broad Balmer lines that can be double-peaked \citep[e.g.,][]{komossa08}. However, TDEs show single sharp outbursts in the X-rays \citep[e.g.,][]{lin11} or the optical--UV \citep[e.g.,][]{vanvelzen21} that then fade away on a timescale of months to years, very different from the long-term light curve of \iras.

There were no major changes in the X-ray hardness ratio in \iras{}  despite strong flux variations over 14 years. This behavior is similar to that of HE\,1136-2304 \citep[][]{zetzl18, kollatschny18}. However, other changing-look AGN show strong X-ray gradient variations, for example, 1ES\,1927+654 \citep{ricci20,laha22}, Mrk\,1018 \citep{husemann16}, or NGC\,3516 \citep{laha22}. Strong changes in the gradient are often attributed to enhanced X-ray absorption by warm-absorber gas. The X-ray spectrum  of IRAS\,23226-2304 was unabsorbed during the minimum state in 2017 as well as during the maximum state in 2019. Therefore, it is unlikely that the changing-look scenario in \iras{} is caused by obscuring clouds. 

The optical luminosity shows the same variability trends as the X-ray luminosity in \iras{}. This is inconsistent with magnetic flux inversion. If magnetic flux inversion would be the dominant effect one would not expect the observed correlated trends \citep{scepi21}. For example, the changing-look AGN 1ES1927+654 showed no correlation between the optical/UV and X-ray emission variations \citep{laha22} in contrast to what we observe for \iras{}.
%{\bf There are examples where the variations are uncorrelated with each other (\citealt{laha22}): e.g., the changing-look AGN 1ES1927+654 showed strong differences in the optical/UV and X-ray emission variations in contrast to that what we observe for \iras{}.
Currently the assumption of a magnetic flux inversion cannot reproduce the behaviour of changing-look AGN in general.

The changing-look character in \iras{} is most probably caused by changes of the accretion rate. This explains the short-term variations on timescales of weeks to months. In addition, the accretion disk structure and/or the distribution and ionization state of the emission-line clouds might have changed. These changes might explain the line profile variations on timescales of years that are independent of the continuum flux variations. Line profile variations independent of intensity variations have been observed as well in the changing-look AGN HE\,1136-2304 \citep{zetzl18,kollatschny18}.
Timescales to explain the changing-look phenomenon have been discussed by, for example, \cite{stern18}. 
%The relevant timescales for changes at the inner accretion 
%disk are either the thermal or the heating/cooling front timescale. 
%The strong, rapid and simultaneous outburst in the continuum
%and the Balmer and FeII lines can not be explained in a
%simple way
%by normal photoionization variations. In that case one would %expect a delay of the FeII lines with respect to the Balmer %lines, and in addition a
%delay of the Balmer lines with respect to the continuum
%variations. Furthermore, a smaller variability amplitude would %have been expected w.r.t. the Balmer line amplitudes.
%In addition, the observed strong variations happened on time %scales of weeks. 

\iras{} shows a low Eddington ratio as discussed in \cite{kollatschny18}. This is consistent with the general trend based on SDSS spectra that the Eddington ratio decreases when the maximum g-band variability increases \citep{rumbaugh18}. Furthermore, \iras{} shows very broad emission lines. This is consistent with the correlation between the strength of optical continuum variability amplitudes in
AGN and their emission line widths \citep{kollatschny06}.

\section{Summary}

We present results on long-term variations and on an outburst  seen in the X-rays, UV, and optical of the changing-look AGN \iras{}. The outburst in 2019 has been observed with \swift{}, XMM-Newton, NuSTAR, SALT, and the SAAO\,1.9\,m telescope. Our findings are summarized below.

\begin{enumerate}[(1)]
\item  The optical, UV, and X-ray continuum light curves of \iras{} showed strong variations with a similar  variability pattern from 1997 until 2021.
There was a strong X-ray, UV,  as well as optical outburst in 2019.  It varied in the X-ray continuum by a factor of 5 and in the optical continuum by a factor of 1.6 within two months. We got a factor of 3 when we corrected for the host galaxy contribution.  The Balmer and \ion{Fe}{ii} emission line intensities showed comparable variability amplitudes for the outburst in 2019.
  
\item The \Ha{} emission line profiles of \iras{} changed from a  blue-peaked profile in the years 1997 and 1999 to a double-peaked profile in the years 2017 and 2019. There were no major variations in the double-peaked profile during the strong outburst in 2019. In 2020, after the outburst, \iras{} changed from a Seyfert type 1 to a Seyfert type 2 object.

\item  A strong outflow component at $v=\,-10\,400$\,\kms{} is to be seen in the Balmer lines. There is additional evidence for an outflow component
based on the deep X-ray spectra.  This absorption feature is present in the Fe band with an outflow velocity of $\sim 0.12c$ ($-36\,000$\,\kms).

\item We took a deep broadband \xmm/\nustar\ spectrum during its maximum state in 2019. This spectrum is qualitatively very similar to a spectrum taken in 2017, but by a factor of 10 more luminous.  The soft X-ray band appears featureless. The soft excess is well modeled with a Comptonization model. A broadband fit with a power-law continuum, Comptonized soft excess, and Galactic absorption gives a good fit to the combined EPIC-pn and NuSTAR spectrum. There is no indication of a warm absorber in the X-ray spectra.
In addition, we see a complex and broadened Fe K emission line profile.

\item The changing-look character in \iras{} is most probably caused by changes in the accretion rate -- based on the short-term variations on timescales of weeks to months. In addition, there are line-profile variations on timescales of years that are independent of the continuum flux variations. These changes might be explained by
changes in the accretion disk structure and/or distribution of the emission line clouds.  

\item The double-peaked Balmer lines are extremely broad. In addition, \iras{} exhibits strong \ion{Fe}{ii} blends. This is in contrast to what is expected from Eigenvector 1 studies.
 \end{enumerate}

\iras{} is an interesting changing-look AGN due to its strong variations in the optical and X-ray continuum, as well as its clear transitions from a single-peaked to a double-peaked Seyfert 1 type object, and finally, to an object of Seyfert type 2. Thus, \iras{} deserves further studies in the future.

\begin{acknowledgements}

This paper is based on observations obtained with XMM-Newton, an ESA science mission with instruments and contributions directly funded by ESA Member States and NASA, as well as observations made with the Southern African Large Telescope (SALT) under programs 2017-1-SCI-030, 2019-1-DDT-003, and 2019-2-DDT-001.\\

This work also made use of data from the NuSTAR mission, a project led by the California Institute of Technology, managed by the Jet Propulsion Laboratory, and funded by the National Aeronautics and Space Administration.
The authors made use of the NuSTAR Data Analysis Software (NuSTARDAS) jointly developed by the ASI Science Data Center (ASDC, Italy) and the California Institute of Technology (USA).\\

%This paper includes observations made with the Southern African Large
%Telescope (SALT) under programs 2017-1-SCI-030, 2019-1-DDT-003, and
%2019-2-DDT-001.\\

This research has made use of the XRT Data Analysis Software (XRTDAS) developed under the responsibility of the ASI Science Data Center (ASDC), Italy, and of the NASA/IPAC Extragalactic Database (NED) which is operated by the Jet Propulsion Laboratory, Caltech, under contract with the National Aeronautics and Space Administration. \\

%The Legacy Surveys consist of three individual
%and complementary projects: the Dark Energy Camera Legacy Survey (DECaLS; Proposal ID
%2014B-0404; PIs: David Schlegel and Arjun Dey), the Beijing-Arizona Sky Survey (BASS;
%NOAO Prop. ID 2015A-0801; PIs: Zhou Xu and Xiaohui Fan), and the Mayall z-band Legacy
%Survey (MzLS; Prop. ID 2016A-0453; PI: Arjun Dey). DECaLS, BASS and MzLS together
%include data obtained, respectively, at the Blanco telescope, Cerro Tololo
%Inter-American Observatory, NSF’s NOIRLab; the Bok telescope, Steward Observatory,
%University of Arizona; and the Mayall telescope, Kitt Peak National Observatory,
%NOIRLab. The Legacy Surveys project is honored to be permitted to conduct astronomical
%research on Iolkam Du’ag (Kitt Peak), a mountain with particular significance to the
%Tohono O’odham Nation.\\

This paper includes a Legacy Survey image. The DESI Legacy Imaging Surveys consist of three individual and complementary projects: the Dark Energy Camera Legacy Survey (DECaLS), the Beijing-Arizona Sky Survey (BASS), and the Mayall z-band Legacy Survey (MzLS). DECaLS, BASS and MzLS together include data obtained, respectively, at the Blanco telescope, Cerro Tololo Inter-American Observatory, NSF’s NOIRLab; the Bok telescope, Steward Observatory, University of Arizona; and the Mayall telescope, Kitt Peak National Observatory, NOIRLab. NOIRLab is operated by the Association of Universities for Research in Astronomy (AURA) under a cooperative agreement with the National Science Foundation. Pipeline processing and analyses of the data were supported by NOIRLab and the Lawrence Berkeley National Laboratory (LBNL). Legacy Surveys also uses data products from the Near-Earth Object Wide-field Infrared Survey Explorer (NEOWISE), a project of the Jet Propulsion Laboratory/California Institute of Technology, funded by the National Aeronautics and Space Administration. Legacy Surveys was supported by: the Director, Office of Science, Office of High Energy Physics of the U.S. Department of Energy; the National Energy Research Scientific Computing Center, a DOE Office of Science User Facility; the U.S. National Science Foundation, Division of Astronomical Sciences; the National Astronomical Observatories of China, the Chinese Academy of Sciences and the Chinese National Natural Science Foundation. LBNL is managed by the Regents of the University of California under contract to the U.S. Department of Energy. The complete acknowledgments can be found at \url{https://www.legacysurvey.org/acknowledgment/}. \\

The authors thank the \swift{} team for performing the ToO observations, and Eileen Herwig for the reduction of the SALT spectra from 2017. ERC acknowledges support from the South African National Research Foundation. This work has been supported by the DFG grant Ko 857/35-1.

\end{acknowledgements}

\bibliographystyle{aa} % style aa.bst
\bibliography{literature} % your references Yourfile.bib

\clearpage

\begin{appendix}
\section{\bf Additional tables}
\begin{table*}
\caption{\label{swiftlog} 
 {XRT and UVOT monitoring observation log: 
Julian date, UT date, and XRT and UVOT exposure times in seconds.}
%Exposure times in s of the \swift{} XRT and UVOT observations
%of \he.
}
%\tabcolsep+6mm
%\centering
%\vspace{3mm}
\begin{tabular*}{\textwidth}{@{\extracolsep{\fill} }ccrrrrrrr}
\hline 
\noalign{\smallskip}
& UT date & \\
\rb{MJD} &  middle of the exposure)  &  \rb{XRT} & \rb{V} & \rb{B} & \rb{U} & \rb{UV W1} & \rb{UV M2} & \rb{UVW2}   \\
% (1) & (2) & (3) & (4) \\ %& (5) & (6) & (7) & (8) & (9) \\ 
%\noalign{\smallskip}
\hline 
%\noalign{\smallskip}
54266.1750 & 2007-06-14 04:12 &  5037 & --- & --- & --- & --- & --- & 4970 \\
54312.4792 & 2007-07-30 11:30 &  6213 & --- & --- & 6127 & --- & --- & --- \\
54361.4743 & 2007-09-17 11:23 & 15767 & --- & --- & --- & --- & --- & --- \\ 
54362.5389 & 2007-09-19 12:56 &  6645 & --- & --- & --- & --- & --- & --- \\ 
54363.6139 & 2007-09-20 14:44 &  8616 & --- & --- & --- & --- & --- & --- \\ 
54676.2875 & 2008-07-29 06:54 &  1095 & 104 & 104 & 104 & 208 & 114 & 417 \\
55092.4722 & 2009-09-18 11:20 & 17422 & --- & --- & --- & --- & --- & --- \\
56493.4166 & 2013-07-20 10:13 &  1461 & 119 & 119 & 119 & 237 & 340 & 475 \\
56555.8861 & 2013-09-20 21:16 &   522 &  44 &  44 &  44 &  87 & 114 & 176 \\
57577.5396 & 2016-07-08 12:57 &  6263 &  84 &  85 &  84 & 168 & 260 & 5509 \\
57855.7061 & 2017-04-12 16:56 &  5819 &  47 &  47 &  47 &  94 & 134 & 188 \\
57855.8778 & 2017-04-12 21:04 &  1141 &  44 &  44 &  44 &  87  & 138 & 176 \\
57864.5097 & 2017-04-21 12:14 &  1111 &  84 &  84 &  84 & 169 & 295 & 337 \\
57880.2952 & 2017-05-07 07:05 &   936 &  78 &  78 &  78 & 154 & 222 & 309 \\
57891.1903 & 2017-05-18 04:34 &   629 &  48 &  48 &  48 &  97 & 180 & 195 \\
57894.6028 & 2017-05-21 14:28 &   946 &  74 &  74 &  74 & 150 & 222 & 301 \\
57896.0243 & 2017-05-23 00:35 &   962 &  80 &  80 &  80 & 158 & 234 & 317 \\
57902.3451 & 2017-05-29 08:17 &   979 &  80 &  80 &  80 & 160 & 245 & 321 \\
57909.0556 & 2017-06-05 01:20 &   202 &  --- &  31 &  31 &  64 & --- &  58 \\
57912.7708 & 2017-06-08 18:30 &   924 &  74 &  74 &  74 & 148 & 223 & 297 \\ 
57915.9201 & 2017-06-11 22:06 &  1898 & 136 & 136 & 136 & 510 & 401 & 544 \\
57922.9986 & 2017-06-18 23:59 &   941 &  26 &  79 &  79 & 158 & 210 & 317 \\
57930.7806 & 2017-06-26 18:44 &   445 &  37 &  37 &  37 &  76 &  90 & 151 \\
57937.6931 & 2017-07-03 16:37 &   799 &  64 &  64 &  64 & 127 & 192 & 257 \\
57944.0639 & 2017-07-10 01:32 &   832 &  82 &  82 &  82 & 163 &  63 & 326 \\
58706.4279 & 2019-08-11 10:16 &   764 &  58 &  58 &  58 & 116 & 211 & 232 \\
58740.6404 & 2019-09-14 15:22 &   762 &  60 &  60 &  60 & 120 & 189 & 240 \\
58746.2726 & 2019-09-20 06:32 &   482 &  41 &  41 &  41 &   82 &  82 & 165 \\
58759.5462 & 2019-10-03 13:06 &   844 &  70 &  70 &  70 &  140 & 179 & 279 \\ 
58761.9466 & 2019-10-05 22:43 &   387 &  --- &  72 & 72 &   144 & ---  &  79 \\
58769.3119 & 2019-10-13 07:29 &  1011 & 81 &  81 &  81 & 163 & 241 & 327 \\
58769.4993 & 2019-10-13 11:59 &    178 & ---  & ---  & --- & ---    & 161 &  \\
58770.7026 & 2019-10-14 16:51 &  1114 & --- & ---  & --- & ---     & --- & 175 \\
58779.6649 & 2019-10-23 15:57 &   487 &  38 &  38 &  38 &   75 & 116 & 151 \\
58782.0028 & 2019-10-26 00:04 &   183 &  --- & --- & 33  & 141 & --- & --- \\
58792.7181 & 2019-11-05 17:14 & 1613 & --- & ---  & 1601 & --- & --- & --- \\
58793.4931 & 2019-11-06 11:50 &   662 &  51 &  51 &  51 & 101 & 172 & 204 \\
58796.5980 & 2019-11-09 14:21 & 1014 &  81 &  81 &  81 & 162 & 250 & 324 \\
58804.5670 & 2019-11-17 13:36 &  889 &  71 &  71 &  71  & 142 & 218 & 283 \\
58823.3669 & 2019-12-06 08:48 &  862 &  71 &  71 &  71  & 143 & 189 & 284 \\ 
58830.8669 & 2019-12-13 20:48 &  994 &  82 &  82 &  82  & 163 & 224 & 328 \\
58837.5791 & 2019-12-20 13:53 &  552 &  69 &  69 &  69  & 138 & ---  & 254 \\  
58851.0457 & 2020-01-03 01:05 & 1044 & 83 &  83 &  83  & 166 & 258 & 333  \\
58858.2817 & 2020-01-10 06:45 & 1064 & 83 &  83 &  83  & 168 & 271 & 336 \\
59312.1285 & 2021-04-08 03:05 &  659 & 51  &  51 &  51 &  102 & 165 & 205 \\
59313.2613 & 2021-04-09 06:16 &  864 & 71  &  71 & 71  &  141 & 195 & 281 \\
%\noalign{\smallskip}
\hline 
%\vspace{-.7cm}
\end{tabular*}
\label{swiftlog}
\end{table*}

\begin{table*}
\caption
{  {\swift{} monitoring: 
MJD, XRT 0.3--10 keV count rates (CR),
and hardness ratios (HR$^1$), X-ray photon index $\Gamma$, the absorption corrected 
0.3--10 keV X-ray flux in units of $10^{-12}$ erg s$^{-1}$ cm$^{-2}$, and
 reduced $\chi^2$ of the simple power-law model fit (pl) with Galactic Absorption.}
}
\tabcolsep+1mm
%\centering
%\vspace{3mm}
\begin{tabular*}{\textwidth}{@{\extracolsep{\fill} } lclcrc}
\hline 
\noalign{\smallskip}
MJD & CR & HR$^1$ &    $\Gamma_{\rm pl}$ & XRT Flux & $(\chi^2/\nu)_{\rm pl}$ \\
% (1) & (2) & (3) & (4) \\ %& (5) & (6) & (7) & (8) & (9) \\ 
%\noalign{\smallskip}
\hline 
%\noalign{\smallskip}
54266.1750  & 0.286\plm0.011  & +0.208\plm0.024  & 1.88\plm0.08 & 11.40\plm0.42 & 39.8/52 \\
54312.4792  & 0.238\plm0.008  & +0.178\plm0.028  & 1.96\plm0.08 & 8.30\plm0.31  & 47.0/49 \\
54361.4743  & 0.083\plm0.003  & +0.266\plm0.033  & 1.78\plm0.09 & 3.10\plm0.14  & 48.7/42 \\ 
54362.5389  & 0.086\plm0.004  & +0.386\plm0.051  & 1.67\plm0.14 & 3.52\plm0.24  & 11.4/18 \\
54363.6139  & 0.079\plm0.003  & +0.255\plm0.044  & 1.89\plm0.11 &  2.81\plm0.09 & 25.9/25  \\
54676.2875  & 0.288\plm0.020  & +0.311\plm0.055  & 1.70\plm0.17 & 11.45\plm0.76 & 167/154$^2$ \\
55092.4722  & 0.340\plm0.007  & +0.138\plm0.015  & 2.04\plm0.04 & 11.46\plm0.02 & 216.0/163$^2$  \\
56492.4166  & 0.073\plm0.008  & +0.305\plm0.096  & 1.82\plm0.29 &  2.91\plm0.35 & 65.1/74$^2$ \\
56555.8861  & 0.149\plm0.018  & +0.241\plm0.127  & 1.92\plm0.39 &  5.09\plm0.92 & 49.5/50$^2$  \\
57577.5396  & 0.051\plm0.003  & +0.345\plm0.056  & 1.77\plm0.16 &  2.10\plm0.24 & 149.4/172$^2$ \\
57855.7061  & 0.037\plm0.009  & +0.212\plm0.240  & 1.78\plm0.47 &  1.65\plm0.33 & 27.7/33$^2$  \\
57855.8778  & 0.045\plm0.010  & +0.367\plm0.210  & ---- &  1.11\plm0.30$^3$ & --- \\
57864.5097  & 0.045\plm0.010  & +0.057\plm0.177  & 2.04\plm0.62 & 1.35\plm0.40 & 19.0/24$^2$ \\
57880.2952  & 0.040\plm0.008  & +0.562\plm0.136  & 2.30\plm1.02 & 2.48\plm0.16 & 18.4/26$^2$ \\
57891.1903  & 0.029\plm0.010  & +0.383\plm0.232  & --- & 1.43\plm0.42$^3$ & ---  \\
57894.6028  & 0.032\plm0.008  & +0.554\plm0.153  & 1.77\plm0.58 & 1.41\plm0.45 & 12.2/22$^2$  \\
57895.0243  & 0.024\plm0.006  & -0.224\plm0.241   & 2.25\plm0.67 & 1.96\plm0.34 & 16.8/16$^2$   \\
57902.3451  & 0.018\plm0.007  & +0.348\plm0.310  & --- & 0.67\plm0.27$^3$ & ---  \\
57909.0556  & 0.017\plm0.013  & ----                      & --- & 0.65\plm0.50$^3$ & ---  \\
57912.7708  & 0.030\plm0.007  & -0.002\plm0.220   & 2.52\plm0.66 & 1.30\plm0.32 & 14.0/16$^2$ \\
57915.9201  & 0.018\plm0.004  & +0.028\plm0.200  & 2.01\plm0.89 & 0.33\plm0.12 & 15.2/28$^2$  \\
57922.9986  & 0.016\plm0.005  & -0.211\plm0.310   &  --- & 0.62\plm0.29$^3$ & ---\\
57930.7806  & 0.014\plm0.007  & ----                      & --- & 0.54\plm0.27$^3$ & ---  \\
57937.6931  & 0.021\plm0.006  & +0.056\plm0.190   & --- & 0.65\plm0.23$^3$ & ---  \\
57944.0639  & 0.022\plm0.006  & +0.435\plm0.221   & --- & 0.77\plm0.23$^3$ & --- \\
58705.4279  & 0.403\plm0.033  & +0.137\plm0.074   & 1.87\plm0.16 & 11.77\plm0.66 & 116/151$^2$ \\
58740.6404  & 0.200\plm0.018  & +0.133\plm0.079   & 1.92\plm0.23 & 6.59\plm0.84 & 87.6/91$^2$ \\
58746.2726  & 0.355\plm0.042  & +0.073\plm0.053   &2.07\plm0.22 & 10.60\plm1.10 & 70.7/101$^2$ \\
58759.5462  & 0.298\plm0.030  & +0.219\plm0.073   & 1.84\plm0.17 & 10.04\plm0.88 & 167/143$^2$\\
58761.9466  & 0.400\plm0.048  & +0.186\plm0.083   & 2.01\plm0.21 & 13.89\plm1.30 & 104/101$^2$ \\
58769.3119  & 0.161\plm0.014  & +0.160\plm0.058   & 1.98\plm0.25 & 3.95\plm0.50 & 74.0/79$^2$ \\
58769.4993  & 0.179\plm0.014  & +0.000\plm0.190   & 1.71\plm0.20 & 7.55\plm0.57 & 116/122$^2$ \\
58770.7026  & 0.148\plm0.013  & +0.056\plm0.094   & 2.05\plm0.25 & 3.37\plm0.48 & 75.0/74$^2$ \\
58779.6649  & 0.303\plm0.028  & +0.257\plm0.093   & 1.95\plm0.23 & 9.28\plm1.15 & 117/81$^2$ \\
58782.0028  & 0.337\plm0.049  & +0.355\plm0.135   & 1.50\plm0.47 & 16.86\plm4.71 & 32.5/38$^2$ \\
58792.7181  & 0.265\plm0.016  & +0.095\plm0.064   & 1.91\plm0.14 &  8.26\plm0.43 & 191/180$^2$ \\
58793.4931  & 0.266\plm0.016  & +0.108\plm0.099   & 1.78\plm0.24 & 15.13\plm1.95 & 96.2/80$^2$ \\
58796.5980  & 0.306\plm0.022  & +0.198\plm0.049    & 1.91\plm0.23 & 9.28\plm1.25 & 117/83$^2$ \\
58804.5670  & 0.154\plm0.015  & +0.091\plm0.076   & 1.98\plm0.25 & 4.91\plm0.70 & 75.2/84$^2$ \\
58823.3669  & 0.127\plm0.015  & +0.062\plm0.125   & 1.90\plm0.31 & 3.57\plm0.49 & 28.9/58$^2$ \\
58830.8669  & 0.152\plm0.014  & +0.362\plm0.081   & 1.69\plm0.22 & 5.22\plm0.42 & 79.3/91$^2$ \\
58837.5791  & 0.291\plm0.033  & +0.070\plm0.087   & 1.90\plm0.23 & 8.70\plm0.71 & 72.4/88$^2$ \\
58851.0457  & 0.233\plm0.018  & +0.247\plm0.072   & 1.69\plm0.17 & 8.49\plm0.82 & 98.6/144$^2$ \\
58858.2817  & 0.308\plm0.023  & +0.168\plm0.071   & 1.93\plm0.17 & 7.81\plm0.73 & 119/133$^2$ \\
59312.1285  & 0.085\plm0.013  & +0.121\plm0.150   & 1.95\plm0.40 & 2.69\plm0.40 & 36.0/36$^2$ \\
59313.2613  & 0.138\plm0.014  & +0.341\plm0.088   & 1.63\plm0.25 & 5.11\plm0.61 & 54.8/79$^2$ \\
%\noalign{\smallskip}
\hline 
%\vspace{-.7cm}
\end{tabular*}
$^1$ The hardness ratio is defined as HR = $\frac{hard-soft}{hard+soft}$ , where {\it \textup{soft}} and {\it \textup{hard}} are the background-corrected counts in the 0.3--1.0 keV and 1.0--10.0 keV bands, respectively

$^2$ Fit using Cash statistics (\citealt{cash79})

$^3$ Extrapolated from the XRT count rates using the flux and count rate of the MJD 57865 observation
\label{swiftdata} 
\end{table*}

\begin{sidewaystable*}
\centering
\tabcolsep1.mm
%\vspace{3cm}
\caption{  {\swift{} monitoring: V, B, U, UVOT W1, M2, and W2 reddening-corrected flux  in units of 10$^{-15}$W\,m$^{-2}$(10$^{-12}$ ergs s$^{-1}$ cm$^{-2}$; columns 2 to 7) and magnitudes in the Vega system (columns 8 to 13).}
}
\begin{tabular}{lcccccc|cccccccc}
\hline
\noalign{\smallskip}
MJD & V & B & U &UVW1  & UVM2 & UVW2  & {V} & {B} & {U} & {UV W1} & {UV
M2} & {UVW2}   \\
\noalign{\smallskip}
(1) & (2) & (3) & (4) & (5) & (6) & (7) & (8) & (9) & (10) & (11) & (12) & (13)
 \\  
\noalign{\smallskip}
\hline
\noalign{\smallskip}
54266.1750 &   ---          &   ---          &     ---      &    ---    &     --- &   9.85\plm0.14  &  ---         &  ---          &  ---          &  ---          &  ---          &  15.06\plm0.05 \\
54312.4792 &   ---          &   ---          &   11.16\plm0.52 &    --- &     --- &     ---      &  ---         &  ---          &  15.10\plm0.05  &  ---          &  ---          &  ---         \\
54361.4743 &   ---          &   ---          &   ---     &    ---       &     --- &     ---      &  ---         &  ---          &  ---          &  ---          &  ---          &  ---         \\
54362.5389 &   ---          &   ---          &   ---     &    ---       &     --- &     ---      &  ---         &  ---          &  ---          &  ---          &  ---          &  ---         \\
54363.6139 &   ---          &   ---          &   ---     &    ---       &     --- &     ---      &  ---         &  ---          &  ---          &  ---          &  ---          &  ---         \\
54676.2875 &  22.44\plm1.05 &  17.32\plm0.81 &   11.80\plm0.55 &   9.57\plm0.54  &    9.92\plm0.65  &   8.42\plm0.55  &  14.90\plm0.05 &  15.54\plm0.05  &  15.04\plm0.05  &  15.08\plm0.06  &  15.03\plm0.07  &  15.23\plm0.07 \\       
55092.4722 &    ---         &   ---          &   ---        &    ---       &      ---       &     ---      &  ---         &  ---          &  ---          &  ---   &  ---          &  ---         \\   
56492.9166 &  17.50\plm0.82 &  12.66\plm0.59 &    6.79\plm0.38 &   4.62\plm0.30  &    3.47\plm0.23  &   3.47\plm0.23  &  15.17\plm0.05 &  15.88\plm0.05  &  15.64\plm0.06  &  15.87\plm0.07  &  16.17\plm0.07  &  16.19\plm0.07 \\   
56555.8861 &  19.73\plm1.30 &  14.40\plm0.81 &    7.72\plm0.51 &   5.61\plm0.42  &    4.70\plm0.45  &   4.58\plm0.39  &  15.04\plm0.07 &  15.74\plm0.06  &  15.50\plm0.07  &  15.66\plm0.08  &  15.84\plm0.10  &  15.89\plm0.09 \\  
57577.5396 &  17.02\plm1.12 &   4.68\plm0.31 &    5.75\plm0.38 &   3.63\plm0.27  &    2.88\plm0.24  &   2.68\plm0.15  &  15.20\plm0.07 &  16.96\plm0.07  &  15.82\plm0.07  &  16.13\plm0.08  &  16.37\plm0.09  &  16.47\plm0.06 \\   
57855.2061 &  17.82\plm1.18 &  11.23\plm0.74 &    5.24\plm0.39 &   2.71\plm0.28  &    2.70\plm0.31  &   2.49\plm0.23  &  15.15\plm0.07 &  16.01\plm0.07  &  15.92\plm0.08  &  16.45\plm0.11  &  16.44\plm0.12  &  16.55\plm0.10 \\
57855.5423 &  17.50\plm1.15 &  12.09\plm0.80 &    5.05\plm0.38 &   2.97\plm0.31  &    2.65\plm0.30  &   2.73\plm0.26  &  15.17\plm0.07 &  15.93\plm0.07  &  15.96\plm0.08  &  16.35\plm0.11  &  16.46\plm0.12  &  16.45\plm0.10 \\   
57864.5097 &  16.26\plm0.91 &  11.76\plm0.55 &    5.10\plm0.33 &   3.19\plm0.24  &    2.58\plm0.22  &   2.78\plm0.21  &  15.25\plm0.06 &  15.96\plm0.05  &  15.95\plm0.07  &  16.27\plm0.08  &  16.49\plm0.09  &  16.43\plm0.08 \\   
57880.2952 &  17.18\plm0.97 &  11.98\plm0.56 &    5.91\plm0.39 &   2.99\plm0.22  &    2.47\plm0.23  &   2.61\plm0.22  &  15.19\plm0.06 &  15.94\plm0.05  &  15.79\plm0.07  &  16.34\plm0.08  &  16.54\plm0.10  &  16.50\plm0.09 \\  
57891.1903 &  17.34\plm1.14 &  12.89\plm0.72 &    5.15\plm0.39 &   3.14\plm0.30  &    2.53\plm0.26  &   2.59\plm0.24  &  15.18\plm0.07 &  15.86\plm0.06  &  15.94\plm0.08  &  16.29\plm0.10  &  16.51\plm0.11  &  16.51\plm0.10 \\  
57894.6028 &  17.82\plm1.00 &  10.82\plm0.61 &    5.19\plm0.34 &   2.31\plm0.22  &    2.51\plm0.24  &   2.63\plm0.22  &  15.15\plm0.06 &  16.05\plm0.06  &  15.93\plm0.07  &  16.62\plm0.10  &  16.52\plm0.10  &  16.49\plm0.09 \\   
57895.5243 &  16.71\plm0.94 &  11.87\plm0.55 &    5.64\plm0.37 &   2.99\plm0.22  &    2.33\plm0.22  &   2.47\plm0.21  &  15.22\plm0.06 &  15.95\plm0.05  &  15.84\plm0.07  &  16.34\plm0.08  &  16.60\plm0.10  &  16.56\plm0.09 \\   
57902.3451 &  17.82\plm1.00 &  11.76\plm0.55 &    5.19\plm0.34 &   2.86\plm0.24  &    2.86\plm0.24  &   2.29\plm0.19  &  15.15\plm0.06 &  15.96\plm0.05  &  15.93\plm0.07  &  16.39\plm0.09  &  16.38\plm0.09  &  16.64\plm0.09 \\  
57909.0556 &    ---         &  12.50\plm0.83 &    5.05\plm0.48 &   2.71\plm0.31  &     ---          &   2.03\plm0.34  &  ---           &  15.89\plm0.07  &  15.96\plm0.10  &  16.45\plm0.12  &  ---             &  16.77\plm0.17 \\ 
57912.7708 &  18.16\plm1.02 &  11.87\plm0.55 &    4.78\plm0.31 &   2.78\plm0.23  &    2.49\plm0.23  &   2.47\plm0.21  &  15.13\plm0.06 &  15.95\plm0.05  &  16.02\plm0.07  &  16.42\plm0.09  &  16.53\plm0.10  &  16.56\plm0.09 \\  
57915.9201 &  15.67\plm0.73 &  11.23\plm0.52 &    5.29\plm0.29 &   2.89\plm0.19  &    2.44\plm0.18  &   2.38\plm0.18  &  15.29\plm0.05 &  16.01\plm0.05  &  15.91\plm0.06  &  16.38\plm0.07  &  16.55\plm0.08  &  16.60\plm0.08 \\  
57922.9986 &  17.50\plm0.98 &  11.65\plm0.54 &    5.24\plm0.34 &   2.86\plm0.21  &    2.49\plm0.23  &   2.49\plm0.21  &  15.17\plm0.06 &  15.97\plm0.05  &  15.92\plm0.07  &  16.39\plm0.08  &  16.53\plm0.10  &  16.55\plm0.09 \\  
57930.7806 &  16.41\plm1.24 &  12.09\plm0.80 &    5.49\plm0.47 &   2.89\plm0.30  &    2.40\plm0.35  &   2.54\plm0.26  &  15.24\plm0.08 &  15.93\plm0.07  &  15.87\plm0.09  &  16.38\plm0.11  &  16.57\plm0.15  &  16.53\plm0.11 \\  
57937.6931 &  17.99\plm1.01 &  12.66\plm0.71 &    4.69\plm0.35 &   2.94\plm0.25  &    2.19\plm0.23  &   2.34\plm0.20  &  15.14\plm0.06 &  15.88\plm0.06  &  16.04\plm0.08  &  16.36\plm0.09  &  16.67\plm0.11  &  16.62\plm0.09 \\  
57944.0639 &  16.56\plm0.93 &  12.20\plm0.57 &    4.78\plm0.31 &   2.89\plm0.21  &    1.98\plm0.37  &   2.68\plm0.20  &  15.23\plm0.06 &  15.92\plm0.05  &  16.02\plm0.07  &  16.38\plm0.08  &  16.78\plm0.19  &  16.47\plm0.08 \\  
58705.9279 &  23.07\plm1.30 &  18.99\plm0.89 &   14.18\plm0.80 &  11.83\plm0.78  &   13.20\plm0.87  &  13.72\plm1.04  &  14.87\plm0.06 &  15.44\plm0.05  &  14.84\plm0.06  &  14.85\plm0.07  &  14.72\plm0.07  &  14.70\plm0.08 \\  
58740.6404 &  20.47\plm1.15 &  15.94\plm0.74 &   12.13\plm0.68 &   8.49\plm0.56  &    9.05\plm0.68  &   9.15\plm0.69  &  15.00\plm0.06 &  15.63\plm0.05  &  15.01\plm0.06  &  15.21\plm0.07  &  15.13\plm0.08  &  15.14\plm0.08 \\  
58746.2726 &  22.03\plm1.45 &  18.99\plm0.10 &   14.18\plm0.80 &  11.50\plm0.87  &   12.96\plm0.98  &  13.72\plm1.04  &  14.92\plm0.07 &  15.44\plm0.06  &  14.84\plm0.06  &  14.88\plm0.08  &  14.74\plm0.08  &  14.70\plm0.08 \\  
58759.5462 &  23.07\plm1.30 &  18.13\plm0.85 &   13.80\plm0.77 &  10.11\plm0.67  &    9.92\plm0.75  &  10.90\plm0.72  &  14.87\plm0.06 &  15.49\plm0.05  &  14.87\plm0.06  &  15.02\plm0.07  &  15.03\plm0.08  &  14.95\plm0.07 \\ 
58761.9466 &    ---         &  17.64\plm0.82 &   13.54\plm0.76 &  11.29\plm0.74  &     ---          &  12.63\plm0.96  &  ---           &  15.52\plm0.05  &  14.89\plm0.06  &  14.90\plm0.07  &  ---             &  14.79\plm0.08 \\ 
58769.3119 &  23.07\plm1.08 &  17.32\plm0.81 &   12.70\plm0.71 &   9.84\plm0.65  &    9.92\plm0.65  &   9.49\plm0.62  &  14.87\plm0.05 &  15.54\plm0.05  &  14.96\plm0.06  &  15.05\plm0.07  &  15.03\plm0.07  &  15.10\plm0.07 \\ 
58769.5091 &    ---         &   ---          &   ---           &    ---          &    8.41\plm0.55  &      ---         &  ---           &  ---            &  ---            &  ---            &  15.20\plm0.07  &  ---           \\
58770.7026 &    ---         &   ---          &   ---           &    ---          &     ---          &    9.76\plm0.83  &  ---           &  ---            &  ---            &  ---            &  ---             &  15.07\plm0.09 \\ 
58779.6649 & 22.23\plm1.47  &  18.81\plm1.08 &   13.42\plm0.88  &  10.99\plm0.73  &   10.88\plm0.83  &  10.12\plm0.87  &  14.91\plm1.07 &  15.45\plm0.06  &  14.90\plm0.07  &  14.93\plm0.07  &  14.93\plm0.08  &  15.03\plm0.09 \\
58782.0028 &    ---         &   ---          &   14.45\plm0.95 &  11.50\plm0.65  &     ---          &     ---         &  ---           &  ---            &  14.82\plm0.07  &  14.88\plm0.06  &  ---             &  ---           \\  
58792.7181 &    ---         &   ---          &   11.69\plm0.43 &    ---          &     ---          &      ---         &  ---           &  ---            &  15.05\plm0.04  &  ---            &  ---             &  ---           \\ 
58793.4931 &  24.16\plm1.36 &  17.16\plm0.96 &   12.47\plm0.70 &   9.75\plm0.64  &   10.58\plm0.70  &  10.03\plm0.66  &  14.82\plm0.06 &  15.55\plm0.06  &  14.98\plm0.06  &  15.06\plm0.07  &  14.96\plm0.07  &  15.04\plm0.07 \\ 
58796.5980 &  22.23\plm1.04 &  17.64\plm0.82 &   12.24\plm0.69 &  10.30\plm0.58  &   10.78\plm0.60  &  10.60\plm0.70  &  14.91\plm0.05 &  15.52\plm0.05  &  15.00\plm0.06  &  15.00\plm0.06  &  14.94\plm0.06  &  14.98\plm0.07 \\  
58804.5670 &  19.91\plm1.12 &  17.00\plm0.79 &   11.16\plm0.63 &   8.81\plm0.49  &    7.32\plm0.48  &   7.40\plm0.49  &  15.03\plm0.06 &  15.56\plm0.05  &  15.10\plm0.06  &  15.17\plm0.06  &  15.36\plm0.07  &  15.37\plm0.07 \\ 
58823.3669 &  21.04\plm1.18 &  16.38\plm0.76 &    9.54\plm0.53 &   7.32\plm0.48  &    7.53\plm0.49  &   6.38\plm0.42  &  14.97\plm0.06 &  15.60\plm0.05  &  15.27\plm0.06  &  15.37\plm0.07  &  15.33\plm0.07  &  15.53\plm0.07 \\ 
58830.8669 &  20.28\plm1.14 &  16.38\plm0.76 &   10.46\plm0.59 &   7.67\plm0.50  &    7.67\plm0.50  &   6.94\plm0.45  &  15.01\plm0.06 &  15.60\plm0.05  &  15.17\plm0.06  &  15.32\plm0.07  &  15.31\plm0.07  &  15.44\plm0.07 \\
58837.5791 &    ---         &  10.43\plm0.58 &    7.44\plm0.49 &   7.39\plm0.48  &     ---          &   7.54\plm0.49  &  ---           &  16.09\plm0.06  &  15.54\plm0.07  &  15.36\plm0.07  &  ---             &  15.35\plm0.07 \\  
58851.0457 &  23.50\plm1.10 &  18.81\plm0.88 &   15.13\plm0.70 &  12.05\plm0.79  &   13.08\plm0.86  &  12.98\plm0.86  &  14.85\plm0.05 &  15.45\plm0.05  &  14.77\plm0.05  &  14.83\plm0.07  &  14.73\plm0.07  &  14.76\plm0.07 \\ 
58858.2817 &  24.38\plm1.14 &  19.16\plm0.89 &   15.84\plm0.74 &  14.22\plm0.94  &   14.48\plm0.95  &  15.04\plm0.99  &  14.81\plm0.05 &  15.43\plm0.05  &  14.72\plm0.05  &  14.65\plm0.07  &  14.62\plm0.07  &  14.60\plm0.07 \\
59312.1285 &  19.54\plm1.29 &  14.14\plm0.79 &    8.62\plm0.57 &   5.26\plm0.40  &    6.26\plm0.47  &   4.98\plm0.37  &  15.05\plm0.07 &  15.76\plm0.06  &  15.38\plm0.07  &  15.73\plm0.08  &  15.53\plm0.08  &  15.80\plm0.08 \\
59313.2613 &  19.37\plm1.09 &  15.08\plm0.70 &    8.08\plm0.45 &   5.98\plm0.39  &    5.40\plm0.41  &   4.62\plm0.35  &  15.06\plm0.06 &  15.69\plm0.05  &  15.45\plm0.06  &  15.59\plm0.07  &  15.69\plm0.08  &  15.88\plm0.08 \\
\noalign{\smallskip}
\hline                                                                          
                                     
\noalign{\smallskip}
%\end{tabular*}
\label{swiftmag}
\end{tabular}
\end{sidewaystable*}

\end{appendix}

%%%%%%%%%%%%%%%%%%%%%%%%%%%%%%%%%%%%%%%%%%%%%%%%%%%%%%%%%%%%%%%%%%%%%%%%%%%%5
\end{document}